\documentclass[sigconf]{acmart}

\renewcommand\footnotetextcopyrightpermission[1]{} %

\usepackage{booktabs} 
\usepackage{multirow}

\setcopyright{rightsretained}

\usepackage{amsmath}

\usepackage{url}
\usepackage{hyperref}

\usepackage{flushend}

\acmDOI{}

\acmISBN{}

\acmConference[MLG'18]{14th International Workshop on
  Mining and Learning with Graphs}{August 2018}{London, UK}
\acmYear{2018}
\copyrightyear{2018}

\begin{document}
\title{Temporal graph-based clustering for historical record
       linkage}
\subtitle{Work-in-progress paper}

\author{Charini Nanayakkara}
\orcid{??}  
\affiliation{%
  \institution{Research School of Computer Science,
               The Australian National University}
  \city{Canberra}
  \state{ACT}
  \postcode{2601}
  \country{Australia}
}
\email{charini.nanayakkara@anu.edu.au}

\author{Peter Christen}
\orcid{0000-0003-3435-2015}
\affiliation{%
  \institution{Research School of Computer Science,
               The Australian National University}
  \city{Canberra}
  \state{ACT}
  \postcode{2601}
  \country{Australia}
}
\email{peter.christen@anu.edu.au}

\author{Thilina Ranbaduge}
\orcid{0000-0001-5405-3704}
\affiliation{%
  \institution{Research School of Computer Science,
               The Australian National University}
  \city{Canberra}
  \state{ACT}
  \postcode{2601}
  \country{Australia}
}
\email{thilina.ranbaduge@anu.edu.au}

\renewcommand{\shortauthors}{C. Nanayakkara et al.}

\begin{abstract}
Research in the social sciences is increasingly based on large and
complex data collections, where individual data sets from different
domains are linked and integrated to allow advanced analytics. A
popular type of data used in such a context are historical censuses,
as well as birth, death, and marriage certificates. Individually, 
such data sets however limit the types of studies that can be
conducted. Specifically, it is impossible to track individuals,
families, or households over time. Once such data sets are linked
and family trees spanning several decades are available it is
possible to, for example, investigate how education, health,
mobility, employment, and social status influence each other and
the lives of people over two or even more generations.
A major challenge is however the accurate linkage of historical data
sets which is due to data quality and commonly also the lack of
ground truth data being available. Unsupervised techniques need to
be employed, which can be based on similarity graphs generated by
comparing individual records. In this paper we present results from
clustering birth records from Scotland where we aim to identify all
births of the same mother and group siblings into clusters. We
extend an existing clustering technique for record linkage by
incorporating temporal constraints that must hold between births by
the same mother, and propose a novel greedy temporal clustering
technique. Experimental results show improvements
over non-temporary approaches, however further work is needed to
obtain links of high quality.
\end{abstract}

%
%



\maketitle


\section{Introduction}

Databases that contain personal information, such as censuses or 
historical civil registries~\cite{Rei02}, generally contain multiple
records describing the same individual (entity) or group of
individuals such as families or households, where each individual
will occur in such databases with different types of
roles~\cite{Chr16,Chr17}. A \emph{baby} is born, then recorded as a
\emph{daughter} or \emph{son} in a census, and later she or he might
marry (as a \emph{bride} or \emph{groom}) and become the
\emph{mother} or \emph{father} of her or his own children. Being
able to link such records across different databases will allow the
reconstruction of whole populations and open a multitude of studies
in the health and social sciences that currently are not feasible
on individual databases~\cite{Blo15,Kum14}.

The process of identifying the sets of records that correspond to
the same individual is known as \emph{record linkage}, \emph{entity
resolution}, or \emph{data matching}~\cite{Chr12}. Record linkage
involves comparing pairs of records to decide if the records of a
pair refer to the same entity (known as a \emph{match}) or to
different entities (a \emph{non-match}). In such a comparison
process generally the similarities between the values of a selected
set of attributes are compared to decide if a pair of records is
similar enough to be classified as a match (if for example the
similarities are above a pre-define threshold value). In many
application domains this simple pair-wise linkage process does
however not provide enough information to identify the
relationships between different individuals~\cite{Chr16,Don15}.

Recently, in contrast to traditional pair-wise record linkage,
\emph{group linkage}~\cite{On07} has received significant attention
because of its applicability of linking groups of individuals, such
as families or households~\cite{Chr17,Fu12}. The identification of
relationships between individuals can enrich data and improve the
quality of data, and thus facilitate more sophisticated analysis of
different socio-economic factors (such as health, wealth, occupation,
and social structure) of large populations~\cite{Fu14,Gru05}.
Studying these issues are important to identify how societies evolve
over time and discover the changes that influenced and contributed
for social evolution~\cite{Don15b}. 

\emph{Historical record linkage} involves the linkage of historical
records, including records from censuses as well as from birth, death,
and marriage certificates, to construct longitudinal data sets about
a population. Over the past two decades researchers working in
different domains have studied the problem of historical record
linkage. 
%
In 1996 Dillon investigated an approach to link census records from
the US and Canada to generate a longitudinal database to examine
changes in household structures~\cite{Dil96}.
The Integrated Public Use Microdata Series (IPUMS, see:
\url{https://www.ipums.org/})
is a large project initiated by the Minnesota Population Centre
(MPC) for linking large demographic data collections. The Life-M
project 
is another example of transforming records from birth, marriage,
and death certificates as well as census records into an
intergenerational longitudinal database~\cite{Bai17}. The project
considers US data from the 19th and 20th centuries and aims
to use birth certificates as a basis for historical record linkage
of large historical databases.


The Digitising Scotland project~\cite{Dib12}, which this work is a
part of, aims to transcribe and link all civil registration events
recorded in Scotland between 1856 and 1973. Around 14 million birth,
11 million death, and 4 million marriage records need to be linked to
create a linked database covering the whole population of Scotland
spanning more than a century to allow researchers in various domains
to conduct studies that are currently impossible to do.

Here we present work-in-progress on a specific step used in
traditional family reconstruction as conducted by demographers and 
historians~\cite{Rei02,Wri73}: the \emph{bundling} (clustering) of
birth records by the same mother to identify siblings. Once 
siblings groups have been identified, they can be linked to census,
marriage, and death records using group linkage
techniques~\cite{Fu14b}. Linked bundles of siblings allow a variety
of studies for example about fertility and mortality and how these
have changed over time~\cite{Rei02}.

\smallskip

\textbf{Contributions}:
In this paper we investigate how clustering techniques for entity
resolution~\cite{Has09c,Sae17} can be employed for bundling birth
records by the same mother, where temporal constraints can be
incorporated to ensure no biologically impossible birth records by
the same mother are linked together. We propose and evaluate a
novel greedy temporal clustering approach, and compare it with a
temporal variation of an existing clustering technique for entity
resolution which has shown to work well in a previous
study~\cite{Sae17}. We conduct an empirical study on a data set
from Scotland which has been extensively linked semi-manually by
domain experts~\cite{Rei02} providing us with ground truth data
to calculate linkage quality. We show that temporal clustering
techniques can outperform the linkage using non-temporal techniques
in terms of linkage quality.


\section{Related Work}

Record linkage has been an active field of research for over half a
century in several research domains. Several recent books and
surveys provide different perspectives of this
area~\cite{Chr12,Don15,Har15,Nau10}.

Classification techniques for record linkage can be categorised into
supervised and unsupervised techniques. Clustering techniques, which
are unsupervised, view record linkage as the problem of how to
identify all records that refer to the same entity and to group these
records into the same cluster. Hassanzadeh et al.~\cite{Has09c}
presented a framework to comparatively evaluate different clustering
techniques for record linkage. Saeedi et al.~\cite{Sae17} recently
proposed a framework to perform clustering for record linkage on a
parallel platform using Apache Flink. Both these frameworks have
implemented and evaluated several clustering approaches. In the evaluation by Saeedi et al.~\cite{Sae17} star clustering (as
described and modified in Section~\ref{sec:starcluster}) was one of
the overall best performing techniques compared to other clustering
techniques. Neither of the two frameworks, however, has considered
temporal constraints.

The linkage of historical data collections with the aim to produce
large temporal linked data sets has recently received increased
attention within the context of population
reconstruction~\cite{Blo15,Kum14}. Such linked population
databases can be an exciting resource in areas such as health,
history, and demography because these databases allow answering
complex questions about temporal changes of a society that so far
have been impossible to address. Most projects in historical record
linkage are challenged by low data quality (due to scanning and
transcription errors of handwritten forms), as well as a lack of
ground truth data (which is difficult and expensive to obtain).
Therefore, research in this area has concentrated on either exploiting
the structure in such data sets (such as households and families)
and developed group linkage methods~\cite{Chr17,Fu14,Fu14b,On07} or
collective techniques~\cite{Chr16}. Alternative approaches explore
the use of limited ground truth data for evaluating linkage
quality~\cite{Ant14a,Bai17}.


\section{Temporal Graph Linkage}
\label{sec:linkage}

Our overall linkage approach
consists of two major phases which we describe in detail in this
section. First we generate an undirected graph based on pair-wise
similarity calculations between individual records (birth
certificates in our case). This is followed by a clustering of
records (nodes) in this graph where we do take temporal
constraints between records into account, as we describe in
Section~\ref{sec:temporal}. In Sections~\ref{sec:starcluster}
and~\ref{sec:tempcluster} we discuss two temporal clustering
approaches, the first based on the extension of an existing
star-based clustering approach~\cite{Has09c,Sae17}, while the
second approach generates clusters in a greedy temporal manner.

For notation we use bold letters for lists, sets and clusters
(upper-case bold letters for lists of sets, lists and clusters), and
normal type letters for numbers and text. Lists are shown with
square and sets with curly brackets, where lists have an order but
sets do not.



\subsection{Similarity Graph Generation}
\label{sec:pairwise}

\begin{figure}[t]
\begin{center}
  \begin{footnotesize}
  \setlength\tabcolsep{2pt}
  \begin{tabular}{ll} \hline
  \multicolumn{2}{l}{\textbf{Algorithm 1: \emph{Pair-wise similarity
    graph generation}}} \\ \hline
  \multicolumn{2}{l}{Input:} \\
  - $\mathbf{R}$: & List of records to be linked \\
  - $\mathbf{A}$: & List of attributes from $\mathbf{R}$ to be
    compared \\
  - $\mathbf{S}$: & List of similarity functions to be applied on
    attributes from $\mathbf{A}$ \\
  - $\mathbf{w}$: & List of weights given to attribute similarities,
    with $|\mathbf{w}| = |\mathbf{S}|$ \\
  - $b, r$ & Number of bands and band size for min-hash based
    LSH blocking \\
  - $s_{min}$: & Minimum similarity for record pairs to be added to
    the generated graph \hspace{2mm} \\ \noalign{\smallskip}
	\multicolumn{2}{l}{Output:} \\
  - $\mathbf{G}$: & Undirected pair-wise similarity graph \\
    \noalign{\medskip}
 \end{tabular}
 \begin{tabular}{ll}
   1:  & $\mathbf{V} = \emptyset$,
         $\mathbf{E} = \emptyset$, $\mathbf{G} = (\mathbf{V},
         \mathbf{E})$ \hspace*{12.5mm} // Initialise empty graph
         \hspace*{5.5mm}\\
   2:  & $\mathbf{L} = \mathbf{MinHashLSHIndexing}(\mathbf{R}, b, r)$
         \hspace{1.0mm} // Generate Min-hash index \\
   3:  & \textbf{for} $\mathbf{l} \in \mathbf{L}$ \textbf{do}: 
         \hspace{25mm} // Loop over all Min-hash blocks \\
   4:  & \hspace{3mm} \textbf{for} $(r_i, r_j): r_i \in \mathbf{l},
         r_j \in \mathbf{l}, r_i.id < r_j.id$ \textbf{do}: \\
   5:  & \hspace{6mm} $\mathbf{s}_{i,j} =
         \mathbf{CompareRecords}(r_i, r_j, \mathbf{A}, \mathbf{S},
         \mathbf{w})$ // Compute similarities \\
   6:  & \hspace{6mm} $s_{i,j} = \mathbf{Normalise}(\mathbf{s}_{i,j},
         \mathbf{w})$ \hspace{14mm} // Normalise the similarity \\
   7:  & \hspace{6mm} \textbf{if} $s_{i,j} \ge s_{min}$
         \textbf{then}: \\
   8:  & \hspace{9mm} $\mathbf{AddNodes}(\mathbf{G}.\mathbf{V},
         \{r_i, r_j\})$ \hspace{11.5mm}// Create two new nodes in
         $\mathbf{G}$\\
   9:  & \hspace{9mm} $\mathbf{AddEdge}(\mathbf{G}.\mathbf{E},
         (r_i, r_j), s_{i,j})$ \hspace{8.5mm}
         // Create an edge in $\mathbf{G}$\\
   10: & \textbf{return} $\mathbf{G}$ \\ \hline
  \end{tabular}
  \end{footnotesize}
\end{center}
\label{algo:algo1}
\end{figure}

The steps involved in the pair-wise similarity calculation phase are
outlined in Algorithm~1. The main input to the algorithm is a list of
records, $\mathbf{R}$, which we aim to link and cluster (in our
case we aim to determine which birth records are by the same
mother). We assume each record has a unique numerical identifier,
$r.id$, and a time-stamp, $r.t$, which in our case is the
registration date of a birth certificate.
We use the list $\mathbf{A}$ of attributes which we will compare
between records using the list of similarity functions $\mathbf{S}$.
These are approximate string matching functions such as Jaro-Winkler
or edit distance~\cite{Chr06a}, or functions specific to the content
of an attribute like a numerical year difference
function~\cite{Chr12}. We also provide a list of weights,
$\mathbf{w}$, to be assigned to the calculated similarities. The
value of the similarity $s_a$ for attribute $a \in \mathbf{A}$
between two records $r_i$ and $r_j$ will be calculated as
$s_a(r_i,r_j) = \mathbf{S}_a(r_i, r_j) \cdot w_a$, where $w_a$ is
the weight for attribute $a \in \mathbf{A}$ and $\mathbf{S}_a$ is
the similarity function used on $a$. The attributes and corresponding
weight values we use in our experiments are shown in
Table~\ref{tab:attr} in Section~\ref{sec:experiments}.

In order to prevent a full pair-wise comparison of each record in
$\mathbf{R}$ with every other record in $\mathbf{R}$ (which has a
complexity of $O(|\mathbf{R}|^2)$), we employ min-hashing based on
locality sensitive hashing (LSH)~\cite{Les14} which requires the two
parameters $b$ (the number of min-hash bands) and $r$ (the band
size). Furthermore, we provide a minimum similarity threshold
$s_{min}$ which determines which record pairs are to be included
in the similarity graph $\mathbf{G}$ being generated.

Algorithm~1 starts by initialising an empty graph, followed by the
generation of the min-hash index $\mathbf{L}$ which consists of
\emph{blocks} of records, $\mathbf{l}$. Each block $\mathbf{l} \in
\mathbf{L}$ contains one or more records from $\mathbf{R}$ that
share the same min-hash value based on the content of the attribute
values in $\mathbf{A}$.
In lines 3 and 4 of the algorithm we loop over these blocks
$\mathbf{l} \in \mathbf{L}$ and generate all unique pairs of records
in each block $\mathbf{l}$. In line 5 we compare the unique record
pairs ($r_i,r_j$) from block $\mathbf{l}$ to calculate a vector of
similarities $\mathbf{s}_{i,j}$. We then normalise these similarities
into $0.0 \le s_{i,j} \le 1.0$ in line 6. If this normalised
similarity is at least the minimum similarity threshold $s_{min}$
then in lines 8 and 9 we insert the two records $r_i$ and $r_j$ as
nodes into the similarity graph $\mathbf{G}$, and we create an
undirected edge between $r_i$ and $r_j$ where the edge attribute
is the normalised similarity $s_{i,j}$.

We finally in line 10 return the generate graph $\mathbf{G}$ which
is used in the second phase of our approach to conduct clustering
of the nodes in this graph. While in the pair-wise similarity
calculation algorithm we do not consider any temporal constraints,
we could add a temporal plausibility calculation step after line 6
and only insert a record pair into $\mathbf{G}$ if the pair is both
similar enough and also temporarily possible, as we describe next.


\begin{figure}[t!]
  \centering
  \includegraphics[width=0.4\textwidth]{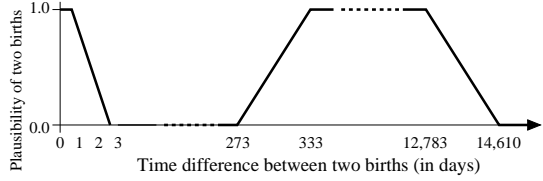}
  \caption{Temporal constraints as the plausibility for the same
    mother to be able to give birth to two children, where the
    horizontal axis shows the time difference (in days) and the
    vertical axis the plausibility $p_{\Delta t}$ that two birth
    records are possible for a certain time difference. Due to
    errors in registration dates, for multiple births we allow for
    a few days difference for twins and triplets, and then have a
    plausible interval between birth from 9 months 
    onwards up-to 35 years. Two births by the same woman more than
    40 years apart is deemed not to be plausible.
    \label{fig:temp-constr}}
\end{figure}

\subsection{Modelling Temporal Constraints}
\label{sec:temporal}

Within the context of clustering birth records by the same mother,
we model temporal constraints as a list $\mathbf{T}$ of time
intervals where it is \emph{plausible} for a mother to have given
birth to two babies.
As illustrated in Figure~\ref{fig:temp-constr}, we need to consider
issues such as data quality as well as multiple births (like twins
and triplets, which potentially are born on two consecutive days).
For each day difference $\Delta t$ between two birth records (i.e.\
the number of days between two births) we calculate a
\emph{plausibility} value $p_{\Delta t}$ (with $0.0 \le  p_{\Delta t}
\le 1.0$), where $p_{\Delta t}=1.0$ for day differences where two
births by the same mother are possible, and $p_{\Delta t}=0.0$ for
day differences where it is biologically not possible for the same
mother to have given birth to two babies. To account for wrongly
recorded dates of birth we apply linear discounting of plausibility
values, as shown in Figure~\ref{fig:temp-constr}.

We can use these temporal plausibility values to modify the similarity
values between records by multiplying normalised record pair
similarities ($s_{i,j}$, as calculated in Algorithm~1) with
plausibility values, and then not considering record pairs in the
graph $\mathbf{G}$ where their new modified similarity is below a
given threshold.

We can apply these temporal constraints during the pair-wise
similarity calculation step described in Section~\ref{sec:pairwise}
(to only include record pairs into the graph $\mathbf{G}$ that are
plausible from a temporal point of view). In the clustering step
described in Sections~\ref{sec:starcluster}
and~\ref{sec:tempcluster} below, we also need to check for
every pair of records in a cluster if they are temporarily plausible.
A cluster can contain pairs of records that are not in $\mathbf{G}$
because their similarity $s_{i,j}$ is below the threshold $s_{min}$,
and these pairs also need to be plausible with regard to the given
temporal constraints. Formally, for a given cluster $\mathbf{c}$, it
must hold: $\forall (r_i \in \mathbf{c}, r_j \in \mathbf{c}):
p_{\Delta t} \ge p_{min}$, where $p_{min}$ is a minimum plausibility
threshold (similar to the similarity threshold $s_{min}$ used in
Algorithm~1). If this condition is not fulfilled for a record $r_i
\in \mathbf{c}$ with all other records in $\mathbf{c}$, then $r_i$
needs to be removed from $\mathbf{c}$. 

While we currently set these temporal intervals of plausible births
by the same mother based on discussions with domain experts, in the
future we aim to learn temporal plausibility values from ground
truth data. Besides temporal constraints between birth records by
the same mother, in our application (where we aim to reconstruct
populations by linking birth, death, marriage, and census records)
there are other constraints we can consider. For example, a death of
an individual can only occur on the same day or after the person's
birth. A marriage should only occur once a person has reached a
minimum age. Similarly, records of the births by a mother can only
occur once she has reached a certain minimum age, and before she has
reached a certain maximum age.


\subsection{Star Clustering}
\label{sec:starcluster}

The second phase of our approach is to use a clustering algorithm
to group all births by the same mother. We selected star clustering
because this algorithm has shown to be one of the best performers in
a previous evaluation study of clustering algorithms for entity
resolution~\cite{Sae17}.
Our contribution to improve star clustering is two-fold: (a) we
introduce temporal constraints as discussed in the previous section,
and (b) we develop several methods for cluster centre selection
and post-processing of overlapping clusters.
Algorithm~2 outlines our modified star clustering algorithm.

\begin{figure}[t]
\begin{center}
  \begin{footnotesize}
  \setlength\tabcolsep{2pt}
  \begin{tabular}{ll} \hline
  \multicolumn{2}{l}{\textbf{Algorithm 2: \emph{Temporal star
    clustering}}} \\ \hline
  \multicolumn{2}{l}{Input:} \\
  - $\mathbf{G}$: & Undirected pair-wise similarity graph \\
  - $\mathbf{T}$: & List of temporal constraints (as discussed in
    Section~\ref{sec:temporal}) \\
  - $p_{min}$: & Minimum plausibility for record pairs to be added
    to a star cluster \\
  - $s_{min}$: & Minimum similarity for record pairs to be
   	added to a star cluster  \hspace{6mm} \\
  - $m_{sort}$: & Method to sort nodes for processing \\
  - $m_{reso}$: & Method to resolve overlapping clusters \\
      \noalign{\smallskip}
	\multicolumn{2}{l}{Output:} \\
  - $\mathbf{C}$: & Final list of clusters \\
    \noalign{\medskip}
 \end{tabular}
 \begin{tabular}{ll}
    1:  & $\mathbf{C} = [\,]$ 
          \hspace{15mm} // Initialise an empty list of clusters \\
    2:  & $\mathbf{U} = [\,]$ \hspace{15mm} // Initialise an empty
          list to hold unassigned nodes \\
    3:  & \textbf{for} $v_i \in \mathbf{G}.V$ \textbf{do}: 
          \hspace{4mm} // Loop over all nodes in graph \\
    4:  & \hspace{3mm} $\mathbf{n}_i =
          \mathbf{GetSimNeighbours}(\mathbf{G}, v_i, s_{min})$
          \hspace{3mm} // Similar neighbours of $v_i$ \\
    5:  & \hspace{3mm} $d_i = |\mathbf{n}_i|$ 
          \hspace{33.5mm} // Degree of $v_i$ \\
    6:  & \hspace{3mm} $a_i =
          \mathbf{CalcAvrSimNeighbours}(\mathbf{G},v_i,
          \mathbf{n}_i)$ \hspace{1mm} // Calculate average
          similarity \\
    7:  & \hspace{3mm} $\mathbf{U}.add((v_i, d_i, \mathbf{n}_i,
         a_i))$ 
         \hspace{1mm} // Add tuple to list of unassigned
         nodes \\
    8:  & $\mathbf{SortTuples}(\mathbf{U}, m_{sort})$ 
         \hspace{6mm} // Sort according to sorting method \\
    9: & \textbf{for} $(v_i, d_i, \mathbf{n}_i, a_i) \in
         \mathbf{U}$ \textbf{do}: \\
   10: & \hspace{3mm} \textbf{U}.$removeTuple$($v_i$) 
         \hspace{2mm} // Remove assigned node from unassigned list \\
   11: & \hspace{3mm} $\mathbf{c}_i = \{v_i\}$ \hspace{8mm}
         // Initialise a new cluster with selected node as centre \\
   12: & \hspace{3mm} \textbf{while} $\mathbf{n}_i \neq \emptyset$ \textbf{do:} \\
   13: & \hspace{6mm} $v_j = $ \textbf{GetNextBestNeighbour($\mathbf{c}_i, \mathbf{n}_i$)}
         \hspace{2mm} // Select next best neighbour \\
   14: & \hspace{6mm} $\mathbf{n}_i.remove(v_j)$
   		 \hspace{2mm} // Remove selected next best neighbour \\
   15: & \hspace{6mm} \textbf{if} 
         \textbf{IsTempPossSimNeighbour}($v_j, \mathbf{c}_i,
         \mathbf{T}, p_{min}$) \textbf{do}:\\
   16: & \hspace{9mm} $\mathbf{c}_i$ $\cup$ $\{v_j\}$
         \hspace{14mm} // Add temporally plausible node to
         cluster \\
   17: & \hspace{9mm} $\mathbf{U}.removeTuple(v_j)$ 
         \hspace{1mm} // Remove node added to the cluster \\
   18: & \hspace{3mm} $\mathbf{C}.add(\mathbf{c}_i)$
         \hspace{19.3mm} // Add cluster to the final cluster list \\
   19: & $\mathbf{v}_{rep} = \mathbf{GetRepeatNodes}(\mathbf{C})$
         \hspace{4.8mm} // Get nodes that occur in multiple
         clusters \\
   20: & $\mathbf{C} = \mathbf{ResolveOverlap}(\mathbf{C},
         \mathbf{v}_{rep}, m_{reso}, s_{min})$ \hspace{1mm}
         // Assign nodes to best cluster \\
   21: & \textbf{return} $\mathbf{C}$ \\ 
        \hline
  \end{tabular}
  \end{footnotesize}
  \vspace*{-3mm}
\end{center}
\label{algo:algo2}
\end{figure}

Our modified algorithm can consider temporal constraints (if the
list of constraints $\mathbf{T}$ is provided) or ignore them (if
$\mathbf{T}$ is empty) when generating clusters. The input to the
algorithm are the pair-wise similarity graph, $\mathbf{G}$, as
generated by Algorithm~1, and the list $\mathbf{T}$ of temporal
constraints. We also require the minimum plausibility $p_{min}$
and minimum similarity $s_{min}$ thresholds to decide if a node
is added to a cluster, and the sorting and overlap resolving
methods, $m_{sort}$ and $m_{reso}$, which we discuss in detail below.

The algorithm starts by initialising an empty list of clusters,
$\mathbf{C}$, and an empty list $\mathbf{U}$ which will hold
information about the nodes that are not yet assigned to clusters.
Initially, all nodes in the similarity graph $\mathbf{G}$ are marked
as unassigned by adding them to $\mathbf{U}$ in the loop starting
in line 3. For each node $v_i \in \mathbf{G}.V$, using the function
$\mathbf{GetSimNeighbours}()$ in line 4 we get the set of its
neighbours $\mathbf{n}_i \in \mathbf{G}$ that have an edge similarity
of at least $s_{min}$. We count the number of these neighbours as the
degree $d_i$ of node $v_i$ in line 5, and also calculate the average
similarity of all edges between $v_i$ and its similar neighbours in
$\mathbf{n}_i$.
In line 7 we append a tuple containing $v_i$, $d_i$, $\mathbf{n}_i$,
and $a_i$ to the list of unassigned nodes $\mathbf{U}$.


Once tuples for all nodes in $\mathbf{G}$ have been added into
$\mathbf{U}$, we sort $\mathbf{U}$ such that the best node to
select as a cluster centre is at the beginning of this list. We
investigate three different methods of how to order nodes based on
the sorting method provided in $m_{sort}$:
%
\setlength{\leftmargini}{3mm}
\begin{itemize}
\item \textbf{Avr-sim-first}: We order the tuples in descending
  order based on their average similarities $a_i$ first and then
  based on the degree $d_i$ (with larger $d_i$ first). With this
  ordering we will process nodes that have high similarities to
  other nodes first.
\item \textbf{Degree-first}: We order the tuples in descending order
  based on their degree $d_i$ first and then based on their average
  similarity $a_i$ (with larger $a_i$ first). With this ordering we
  will process nodes that have many edges with high similarities
  to other nodes first.
\item \textbf{Comb}: With this method we order nodes in descending
  order based on combined score where we multiply their average
  similarity with the logarithm of their degree, i.e. $a_i \times
  log(d_i$). We take the logarithm of $d_i$ because $a_i$ is
  normalised into $0 \le a_i \le 1$ while $d_i$ is a positive
  integer value and therefore would dominate the combined score.
  With this method we aim to weigh both degree and average
  similarities to obtain an improved ordering.
\end{itemize}
In lines 9 to 18 of the algorithm, we process one tuple in
$\mathbf{U}$ after another. Only an unassigned node can become the
centre of a new star cluster. The tuple of node $v_i \in \mathbf{U}$
selected to become a star centre is removed from the list of
unassigned nodes and a new cluster $\mathbf{c}_i$ is created in line
11. 
Then we find the next best node to add to cluster $\mathbf{c}_i$,
using the function $\mathbf{GetNextBestNeighbour()}$. This function
selects the node $v_j \in \mathbf{n}_i$ which has the highest
average similarity with the nodes that are currently assigned to the
cluster $\mathbf{c}_i$. The selected node $v_j$ is removed from
$\mathbf{n}_i$ in line 14 so it cannot be selected as the best
neighbour in the next iteration. For each next best neighbour
$v_j$ we check in line 15 if $v_j$ is plausible with every other
node in $\mathbf{c}_i$ with regard to the temporal constraints
given in the list $\mathbf{T}$ using 
$\mathbf{IsTempPossSimNeighour}()$ (if $\mathbf{T}$ is empty then
this function returns true), and the minimum
plausibility threshold $p_{min}$. We add the plausible nodes $v_j$
to the cluster $\mathbf{c}_i$ in line 16 and remove their
corresponding tuples from $\mathbf{U}$ in line 17. This means these
nodes cannot become the centre of another star cluster.



The final steps of Algorithm~2, lines 19 and 20, deal with those
nodes that are members of more than one cluster (note these are
not star cluster centres). Overlapping clusters are not desirable
for record linkage because each cluster represents one entity. In
line 19 we therefore identify the set $\mathbf{v}_{rep}$ of nodes
which occur in more than one cluster in the list $\mathbf{C}$,
and in line 20 we use the function $\mathbf{ResolveOverlap}()$ to
resolve overlapping clusters, where the method $m_{reso}$ determines
how we assign a node $v_j \in \mathbf{v}_{rep}$ to its best cluster.
We investigate three methods to resolve overlaps:
\begin{itemize}
\item \textbf{Avr-all}: We average the
  similarities between the node $v_j$ and all the nodes in a
  cluster it is connected to in the similarity graph $\mathbf{G}$
  by dividing this similarity sum by $n-1$ where $n$ is the number
  of nodes in the cluster (including $v_j$), i.e.\ we do take nodes
  in a cluster which are not connected to $v_j$ in $\mathbf{G}$ into
  account.
\item \textbf{Avr-high}: We calculate the average similarity between
  the node $v_j$ and all the nodes in a cluster it is connected to
  in the similarity graph $\mathbf{G}$, with similarities of at
  least $s_{min}$.
\item \textbf{Edge-ratio}: In this method we count the number of
  edges between $v_j$ and nodes in a cluster that have a similarity
  of at least $s_{min}$ and divide this number by $n-1$ where $n$
  is the number of nodes in the cluster (including $v_j$).
\end{itemize}
For each node $v_j \in \mathbf{v}_{rep}$, we assign it to the
cluster with the highest value according to the selected method to
resolve overlaps. For all three methods, if for a given node $v_j$
two or more clusters have the same calculated score then we assign
$v_j$ to the cluster where $v_j$ has the highest number of similar
edges to. 
At the end of this process, the final list of clusters
$\mathbf{C}$ contains no overlapping clusters.




\subsection{Greedy Temporal Clustering}
\label{sec:tempcluster}

\begin{figure}[t!]
  \centering
  \includegraphics[width=0.36\textwidth]
    {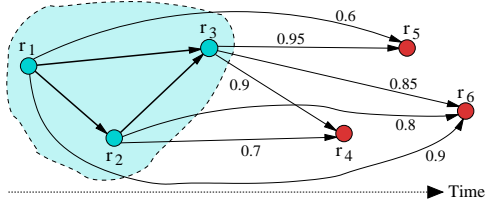}
  \caption{Example of the greedy temporal linkage approach described
    in Section~\ref{sec:tempcluster}, showing nodes (records) and
    edges (similarities) from the directed similarity graph
    $\mathbf{G}_D$. Records $r_1$ to $r_3$ show an existing cluster,
    and the question now is which best future record (from $r_4$,
    $r_5$, and $r_6$) is to be added to the cluster next. We consider
    three selection methods: (a) the earliest next possible
    (according to temporal constraints) record in the graph
    $\mathbf{G}$ (in this example $r_4$), (b) the future record with
    the highest maximum similarity ($r_5$), or (c) the future record
    with the highest average similarity ($r_6$).
    \label{fig:temp-cluster}}
\end{figure}

The second temporal clustering approach is based on the idea of
iteratively adding nodes to clusters using a greedy selection method,
as illustrated in Figure~\ref{fig:temp-cluster}. We initially create
one cluster per record, and insert these singleton clusters into a
priority queue that is sorted according to time-stamps (i.e.\ the
dates of birth registrations in our case) with the smallest
time-stamp first. We then process the earliest cluster first, and aim
to expand this cluster with a new record that is in the future (of
the latest record in the cluster), as Figure~\ref{fig:temp-cluster}
shows. In this greedy approach the question is how to select the best
future node (record) to add to a cluster. We implement (and evaluate
in Section~\ref{sec:experiments}) three different such selection
methods:
\begin{itemize}
\item \textbf{Next}: Select the temporal next record (with the
  smallest time-stamp) that is connected via an edge in the graph
  $\mathbf{G}$ to any record in the cluster. This method does
  neither consider the similarities between nodes (besides the
  edges in $\mathbf{G}$) nor their connectivities and serves as a
  greedy baseline.
\item \textbf{Max-sim}: Select the record in the future that is
  connected via an edge in the graph $\mathbf{G}$ to any record in
  the cluster and that has the highest similarity $s_{i,j}$ with any
  record in the cluster. This method generates clusters where nodes
  are connected via edges of high similarities, however, these
  clusters might not be dense.
\item \textbf{Avr-sim}: Select the record in the future that is
  connected via an edge in the graph $\mathbf{G}$ to one or more
  records in the cluster and that has the highest average
  similarity over these edges. This method generates dense clusters
  with high similarity edges.
\end{itemize}

\begin{figure}[t!]
\begin{center}
  \begin{footnotesize}
  \setlength\tabcolsep{2pt}
  \begin{tabular}{ll} \hline
  \multicolumn{2}{l}{\textbf{Algorithm 3: \emph{Greedy temporal
    clustering}}} \\ \hline
  \multicolumn{2}{l}{Input:} \\
  - $\mathbf{G}$: & Undirected pair-wise similarity graph \\
  - $\mathbf{T}$: & List of temporal constraints (as discussed in
    Section~\ref{sec:temporal}) \\
  - $p_{min}$: & Minimum plausibility for record pairs to be
    considered \hspace{17mm} \\
  - $m_{sel}$: & Method on how to select the next node to add to a
    cluster \\ \noalign{\smallskip}
	\multicolumn{2}{l}{Output:} \\
  - $\mathbf{C}$: & Final list of clusters \\
    \noalign{\medskip}
 \end{tabular}
 \begin{tabular}{ll}
	1:  & $\mathbf{G}_D =
        \mathbf{GenerateTempDirGraph(G)}$ \hspace{2.5mm}
        // A temporal directed graph \\
    2:  & $\mathbf{C} = [\,]$ \hspace{32mm} // Initialise an empty
          list of clusters \\
    3:  & $\mathbf{Q} = [\,]$ \hspace{32mm} // Initialise an empty
          priority queue \\
    4:  & \textbf{for} $v \in \mathbf{G}_D.V$ \textbf{do}:
          \hspace{26mm} // Loop over all nodes in $\mathbf{G}_D$ \\
    5:  & \hspace{3mm} \textbf{if} $(|v.in()| = 0) \wedge
          (|v.out()| = 0)$ \textbf{then}: // A singleton \\
    6:  & \hspace{6mm} $\mathbf{C}.add(\{v\})$ 
          \hspace{25.5mm} // Add to the final list of clusters \\
    7:  & \hspace{3mm} \textbf{else:} \\
    8:  & \hspace{6mm} $\mathbf{Q}.add((v.t, \{v\}))$ 
          // Add node with its time-stamp to queue $\mathbf{Q}$ \\
    9:  & $\mathbf{Sort(Q)}$ \hspace{15mm}
          // Sort queue according to time-stamps (earliest first) \\
    10: & \textbf{while} $\mathbf{Q} \ne []$ \textbf{do}:
          \hspace{5.5mm} // Loop over temporal clusters until
          $\mathbf{Q}$ is empty \\
    11: & \hspace{3mm} $(t, \mathbf{c}_{tmp}) = \mathbf{Q}.pop()$ 
          \hspace{6.5mm} // Get first cluster tuple in $\mathbf{Q}$\\
    12: & \hspace{3mm} $\mathbf{o} = \cup v_i.out(), v_i \in
          \mathbf{c}_{tmp}$ \hspace{0.5mm} // Set of all outgoing
          nodes \\
    13: & \hspace{3mm} \textbf{if} $\mathbf{o} = \emptyset$
          \textbf{do}: 
          \hspace{16mm} // No outgoing nodes found in
          $\mathbf{c}_{tmp}$ \\
    14: & \hspace{6mm} $\mathbf{C}.add(\mathbf{c}_{tmp})$
          \hspace{10.5mm} // Add $\mathbf{c}_{tmp}$ to the final
          list of clusters \\
    15: & \hspace{3mm} \textbf{else}: \\
    16: & \hspace{6mm} \textbf{if} $m_{sel} = $ \textbf{Next}
          \textbf{do}: \hspace{5mm}// Select node with smallest
          time-stamp \\
    17: & \hspace{9mm} $v_n = v_i \in \mathbf{o}: \text{argmin}
          \{ v_i.t: v_i \in \mathbf{o} \}$ \\
    18: & \hspace{6mm} \textbf{if} $m_{sel} = $ \textbf{Max-sim}
          \textbf{do}: \hspace{0.5mm} // Select node with the
          highest similarity \\
    19: & \hspace{9mm} $v_n = v_i \in \mathbf{o}: \text{argmax}
          \{ s_{i,j}: v_i \in \mathbf{c}_{tmp}, v_j \in
          \mathbf{o} \}$ \\
    20: & \hspace{6mm} \textbf{if} $m_{sel} = $ \textbf{Avr-sim}
          \textbf{do}: \hspace{1.5mm} // Select node with highest
          average similarity \\
    21: & \hspace{9mm} $v_n = v_i \in \mathbf{o}: \text{argmax}
          \{\sum s_{i,j}  / |\{(v_i,v_j): v_i \in \mathbf{c}_{tmp},
          v_j \in \mathbf{o} \}| \}$ \\
    22: & \hspace{6mm} $p_{\Delta t} =
          \mathbf{CheckTempConstr}(v_n.t, \mathbf{c}_{tmp},
          \mathbf{T})$ // Temporal plausibility \\
    23: & \hspace{6mm} \textbf{if} $p_{\Delta t} \ge p_{min}$
          \textbf{do}: \\
    24: & \hspace{9mm} $\mathbf{Q}.add((v_n.t, \mathbf{c}_{tmp}
          \cup \{v_n\}))$ 
          \hspace{9mm}// Add expanded
          $\mathbf{c}_{tmp}$ to $\mathbf{Q}$ \\
    25: & \hspace{9mm} $\mathbf{Sort(Q)}$ \hspace{7.5mm} // Sort
          queue according to time-stamps (earliest first) \\
    26: & \hspace{6mm} \textbf{else}: \\
    27: & \hspace{9mm} $\mathbf{C}.add(\mathbf{c}_{tmp})$
          \hspace{1mm} // Add $\mathbf{c}_{tmp}$ to the list of
          final clusters \\
    28: & \textbf{return} $\mathbf{C}$ \\
          \hline
  \end{tabular}
  \end{footnotesize}
\end{center}
\label{algo:algo3}
\end{figure}

As with star clustering, we can consider temporal constraints when
selecting the next record to be added into a cluster, or we can
ignore any temporal constraints. Algorithm~3 outlines the steps
involved in this temporal greedy clustering approach.  

The main input to the algorithm are the pair-wise similarity graph,
$\mathbf{G}$, and a list of temporal constraints, $\mathbf{T}$, as
discussed in Section~\ref{sec:temporal}. We also input a minimum
plausibility threshold $p_{min}$ which is used to consider which
record pairs are to be added into clusters based on their temporal
constraints, and the selection method $m_{sel}$ which determines
which nodes (records) to add into a cluster. 

We first (in line 1) convert the undirected similarity graph
$\mathbf{G}$ into a directed graph where each node (birth record)
has an outgoing edge to any future node, as shown in
Figure~\ref{fig:temp-cluster}. The function
$\mathbf{GenerateTempDirGraph}()$ generates a directed graph
$\mathbf{G}_D$ by considering the time differences between the pairs
of nodes in $\mathbf{G}$, such that $\forall (v_i, v_j) \in
\mathbf{G}_D.E: v_j.t \ge v_i.t$. In line 4, the algorithm then loops
over each node $v \in \mathbf{G}_D$ and adds $v$ to the final list
of clusters $\mathbf{C}$ if $v$ does not have any incoming or
outgoing edges to other nodes (lines 5 and 6), i.e.\ the node is a
singleton. Otherwise, a new cluster is created containing only node
$v$, and this cluster is added together with its time-stamp, $v.t$,
as a tuple into the priority queue $\mathbf{Q}$ for further
processing (line 8).

In line 9 we sort $\mathbf{Q}$ according to the time-stamps of each
cluster such that the cluster with the smallest time-stamp is at
the beginning of the queue. The main loop of the algorithm starts in
line 10 where in each iteration we retrieve the cluster
$\mathbf{c}_{tmp}$ with the earliest time-stamp $t$ (line 11). We
then find for each node $v_c \in \mathbf{c}_{tmp}$ all its outgoing
nodes in $\mathbf{G}_D$, and in line 12 we combine these into the
set $\mathbf{o}$ of all outgoing nodes for $\mathbf{c}_{tmp}$. If
$\mathbf{o}$ is empty for the current cluster $\mathbf{c}_{tmp}$
then $\mathbf{c}_{tmp}$ is added to the final list of clusters
$\mathbf{C}$ in line 14 because it cannot be expanded further.

\begin{table*}[!t]
  \centering
   \begin{small}
  \caption{Attributes in birth certificates used for three variations
    of calculating pair-wise similarities to generate the graph
    $\mathbf{G}$.}
    \label{tab:attr}
  \end{small}
  \begin{footnotesize}
    \begin{tabular}{lccccc}
    \hline\noalign{\smallskip}
    Attribute & Similarity function & Weight & ~All attributes~ &
      ~Parent names only~ & ~Parent names and addresses~ \\
      \noalign{\smallskip}\hline\noalign{\smallskip}
      Father first name & Jaro-Winkler & 6.578 & \checkmark &
        \checkmark & \checkmark \\
      Father last name & Jaro-Winkler & 7.168 & \checkmark &
        \checkmark & \checkmark \\
      Mother first name & Jaro-Winkler & 4.483 & \checkmark &
        \checkmark & \checkmark \\
      Mother last name & Jaro-Winkler & 7.168 & \checkmark &
        \checkmark & \checkmark \\
      Mother maiden last name~ & Jaro-Winkler & 5.985 & \checkmark &
        \checkmark & \checkmark \\ \noalign{\smallskip}
      Parents marriage day & Exact & 4.610 & \checkmark &  & \\
      Parents marriage month & Exact & 3.855 & \checkmark &  & \\
      Parents marriage year & Year difference & 5.240 & \checkmark &
        & \\
      Parents marriage place 1 & Jaro-Winkler & 4.435 & \checkmark &
        & \\
      Parents marriage place 2 & Jaro-Winkler & 3.607 & \checkmark &
        & \\ \noalign{\smallskip}
      Occupation father & Jaro-Winkler & 2.247 & \checkmark &  & \\
      Occupation mother & Jaro-Winkler & 1.274 & \checkmark &  & \\
        \noalign{\smallskip}
      Address 1  & Jaro-Winkler & 4.715 & \checkmark &  &
        \checkmark \\
      Address 2  & Jaro-Winkler & 3.548 & \checkmark &  &
        \checkmark \\
      Source parish & Jaro-Winkler & 4.562 & \checkmark &  & 
        \checkmark \\
      \noalign{\smallskip} \hline
  \end{tabular}
  \end{footnotesize}
\end{table*}

\begin{table}[th]
\centering
   \begin{small}
  \caption{The ten most frequent values and their corresponding
    frequency counts for first and last names of fathers and
    mothers in the Isle of Skye birth data set.}
    \label{tab:freq-values}
  \end{small}
 \begin{footnotesize}
\begin{tabular}{cc cc}
\hline\noalign{\smallskip}
\multicolumn{2}{c}{First name} & \multicolumn{2}{c}{Last name} \\
Father & Mother & Father & Mother \\
\noalign{\smallskip}\hline\noalign{\smallskip}
	John (3,444) & Mary (2,740)  & Mcleod (1,571)    & Mcdonald (1,793) \\
	Donald (2,628) & Catherine (2,607) & Mcdonald (1,556) & Mcleod (1,761) \\
	Alexander (1,665) & Ann (2,084)     & Mckinnon (1,168) & Mckinnon (1,164) \\
	Malcolm (800)    & Margaret (2,031)  & Nicolson (1,047)  & Nicolson (908) \\
	Neil (787)       & Christina (1,626) & Mclean (908)   & Mclean (850) \\
	Angus (782)      & Marion (1,532)    & Campbell (685)   & Campbell (823) \\
	William (611)    & Flora (1,150)     & Mcinnes (682)    & Mcinnes (704) \\
	Murdo (565)      & Janet (871)      & Mckenzie (637)   & Matheson (541) \\
	Norman (513)     & Effie (654)      & Mcpherson (525)  & Mckenzie (509) \\
	Ewen (502)     & Isabella (478) & Robertson (452) & Mcpherson (496) \\
\noalign{\smallskip} \hline
\end{tabular}
 \end{footnotesize}
\end{table}

On the other hand, if there are outgoing nodes (i.e.\ $\mathbf{o}$
is not empty), then based on the selection method $m_{sel}$, as
explained above, the algorithm selects the next best node, $v_n$, to
be added into the current cluster $\mathbf{c}_{tmp}$ in lines 16 to
21. Using the function \textbf{CheckTempConstr()} in line 22 we then
check the temporal plausibility $p_{\Delta t}$ between node $v_n$
and all nodes in $\mathbf{c}_{tmp}$ based on the list of
temporal constraints $\mathbf{T}$ (if this list is empty, i.e.\ no
temporal constraints are given, then we set $p_{\Delta t}=1$). If
the calculated $p_{\Delta t}$ is at least $p_{min}$ (i.e.\ $v_n$ is
temporary plausible with all other nodes in $\mathbf{c}_{tmp}$),
then $v_n$ is added to the current cluster $\mathbf{c}_{tmp}$ and
the expanded cluster is added as a new tuple into $\mathbf{Q}$ with
$v_n.t$ as the tuple's time-stamp (line 24). $\mathbf{Q}$ is sorted
again in line 25 to ensure the cluster with the smallest time-stamp
(of its temporarily last record) is selected in the next iteration
(line 25). If $v_n$ is not temporally plausible with at least one
node in $\mathbf{c}_{tmp}$ then $\mathbf{c}_{tmp}$ is added to the
final list of clusters $\mathbf{C}$ in line 27 because it cannot be
expanded further.


\section{Experimental Evaluation}
\label{sec:experiments}

We evaluate our proposed temporal clustering approaches using a
real Scottish birth data set that covers the population of the
Isle of Skye over the period from 1861 to 1901. This data set
contains 17,614 birth certificates, where each of these contains
personal information about the baby and its parents, as shown in
Table~\ref{tab:attr}.

This data set has been extensively curated and linked semi-manu-ally
by demographers who are experts in the domain of linking such
historical data~\cite{New11,Rei02}. Their approach followed long
established rules for family reconstruction~\cite{Wri73}, leading
to a set of linked birth certificates. We thus have a set of manually
generated links that allows us to compare the quality and coverage
of our automatically identified links to those identified by the
domain experts.

\begin{figure*}[t!]
  \centering
    \includegraphics[width=0.3\textwidth]
      {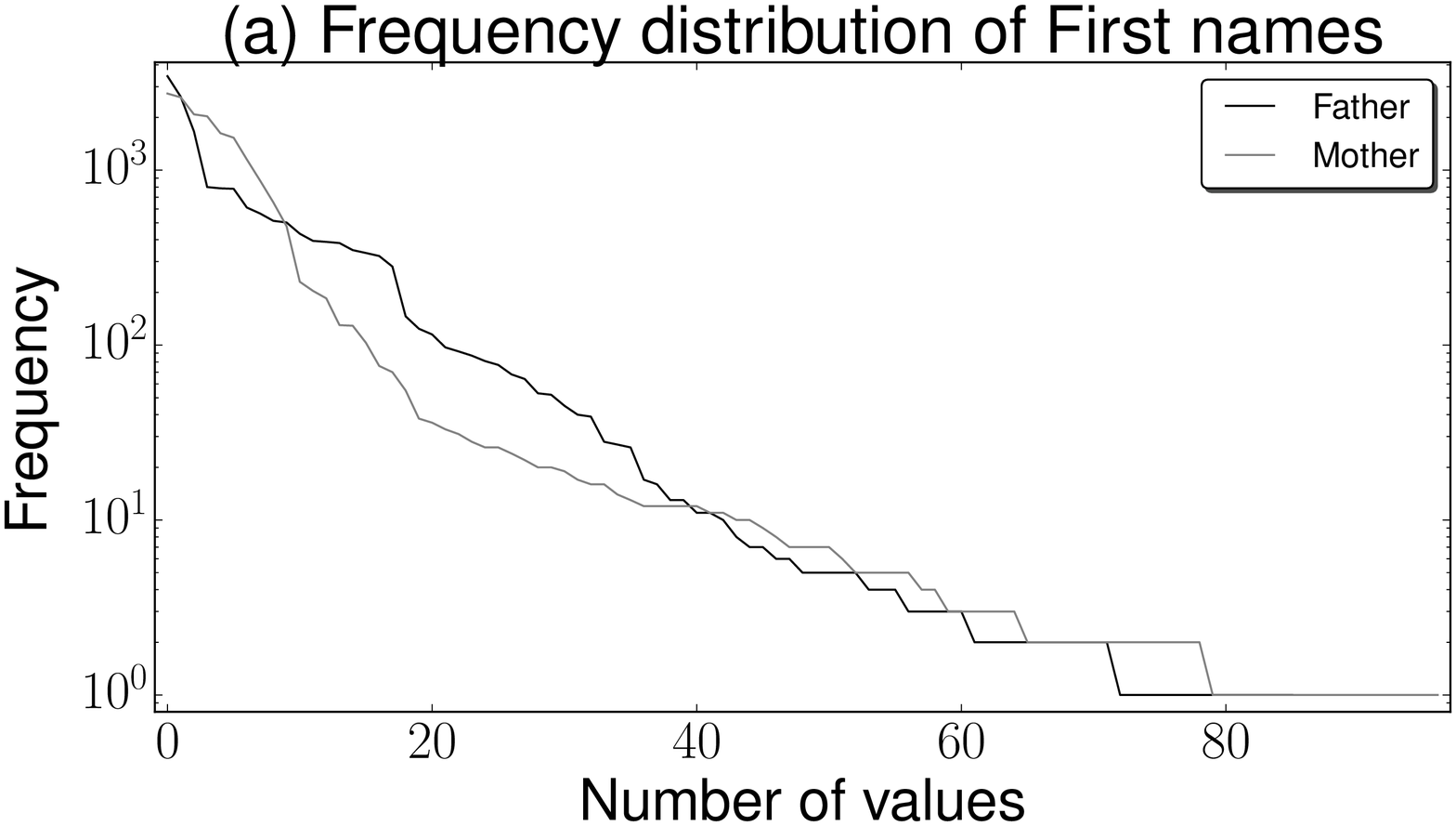}
    \hspace{4mm}
    \includegraphics[width=0.3\textwidth]
      {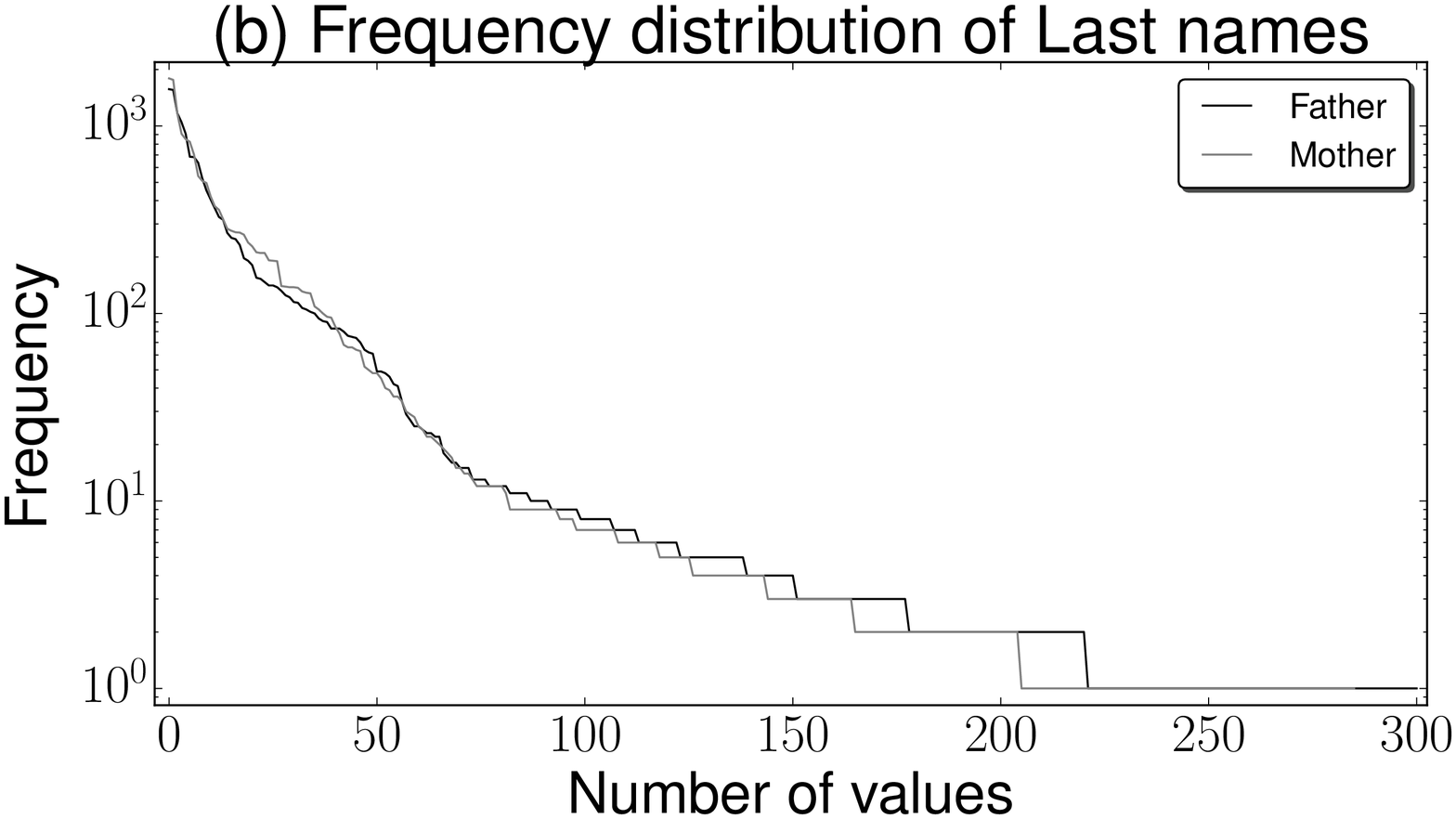}
    \hspace{4mm}
    \includegraphics[width=0.3\textwidth]
      {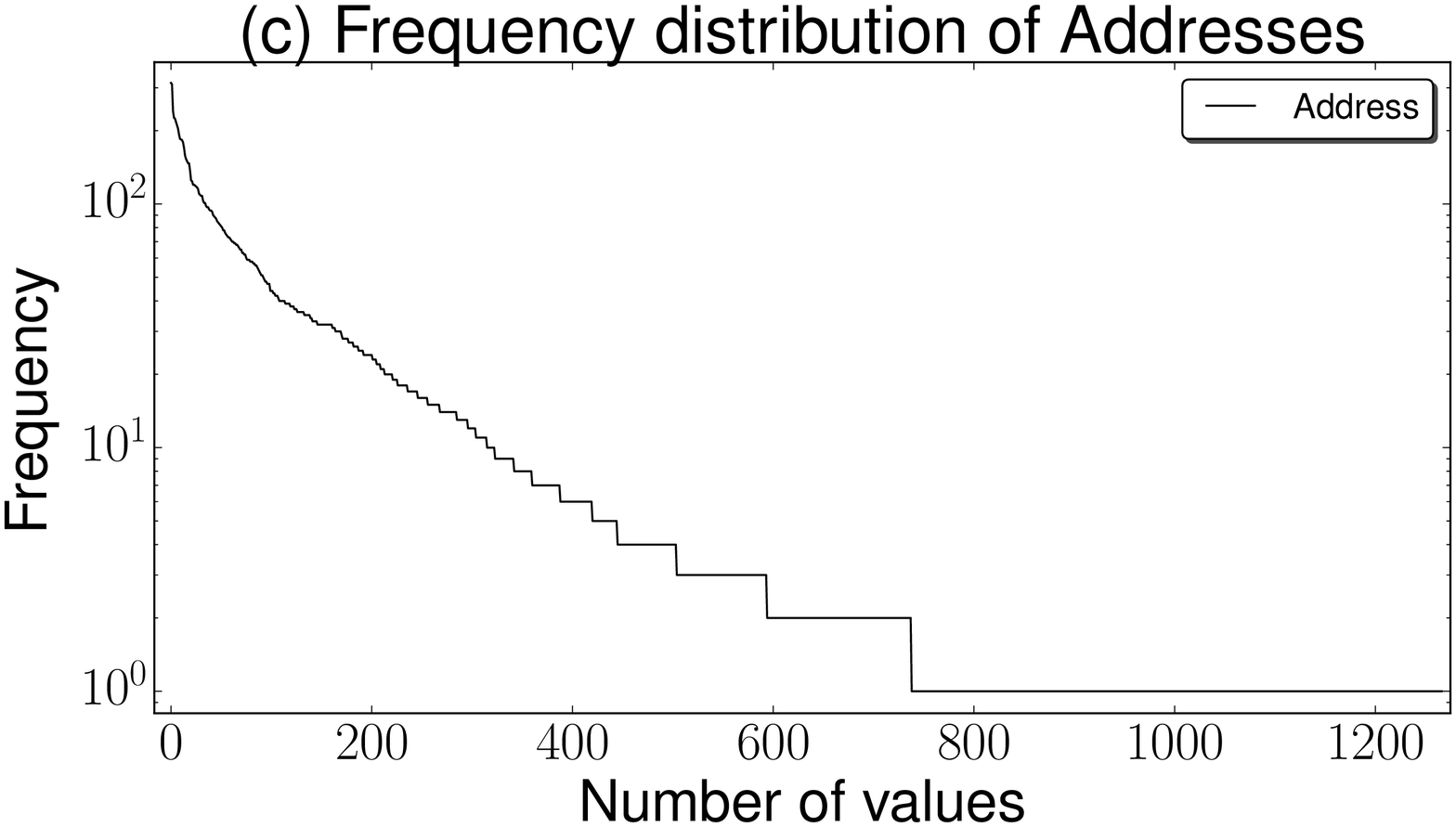}
  \caption{Frequency distribution of (a) first names and (b) last
    names of parents, and (c) addresses in the Isle of Skye birth
    data set. Note the y-axis are in log scale. Notice the highly
    skewed frequency distributions where a few names occur many
    times.     
    \label{fig:freq-dist}}
\end{figure*}

As with other historical data sets~\cite{Ant14a,Fu14b}, this
birth data set has a very small number of unique name values (2,055
first names and only 547 last names). As Figure~\ref{fig:freq-dist}
shows, the frequency distributions of names are also very skewed.
The ten most common first and last name values occur in between
$30\%$ and $40\%$ of all records, as Table~\ref{tab:freq-values}
illustrates. Many records have missing values in address or
occupation attributes, and for unmarried women the details of a
baby's father are mostly missing.

As commonly performed in record linkage research~\cite{Chr12,Nau10},
we evaluate our clustering approaches with regard to precision (how
many of the identified links between birth records are true links
according to the demographers) and recall (how many true links have
our clustering approaches correctly identified and inserted into the same
clusters). We do not present F-measure results given recent work
has identified some problematic aspects when using the F-measure to
compare record linkage approaches~\cite{Han18}.

We implemented all techniques using Python 2.7.6 and used the
string matching functionalities provided in
\emph{Febrl}~\cite{Chr08e} to conduct the pair-wise record
comparisons. We set the LSH min-hash parameters as $b=100$ (number
of bands) and $r=4$ (band size) in order to obtain a recall of
$99.7\%$ of the true matches in the ground truth data set for the
similarity graph $\mathbf{G}$.
We used three different subset of attributes, $\mathbf{A}$, as
described in Algorithm~1 and illustrated in Table~\ref{tab:attr}.
For details of the similarity functions used see~\cite{Chr12}.
%
We calculated attribute similarities with either the weights shown
in Table~\ref{tab:attr}, or with all attribute weights set to $1.0$.
We thus ended up with six similarity graphs where we set
$s_{min}=0.7$: \emph{weighted} and \emph{no weights}, and \emph{All
attributes}, \emph{Parent names and addresses}, and \emph{Parent
names only}. This allows us to investigate how different ways to
calculate pair-wise similarities influence the quality of the final
clustering.

\begin{figure*}[t!]
  \centering

    \includegraphics[width=0.3\textwidth]
      {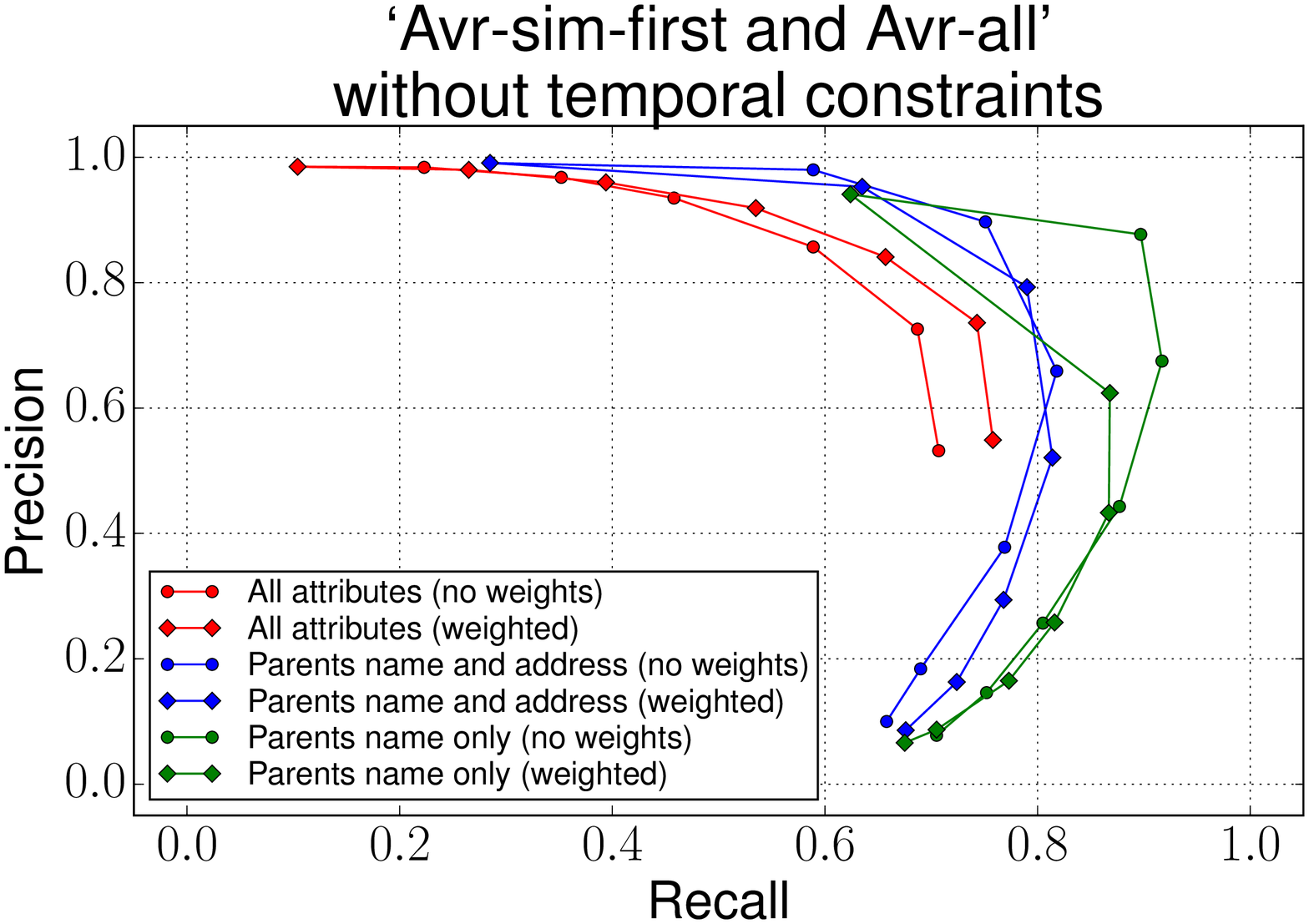}
    \hspace{4mm}
    \includegraphics[width=0.3\textwidth]
      {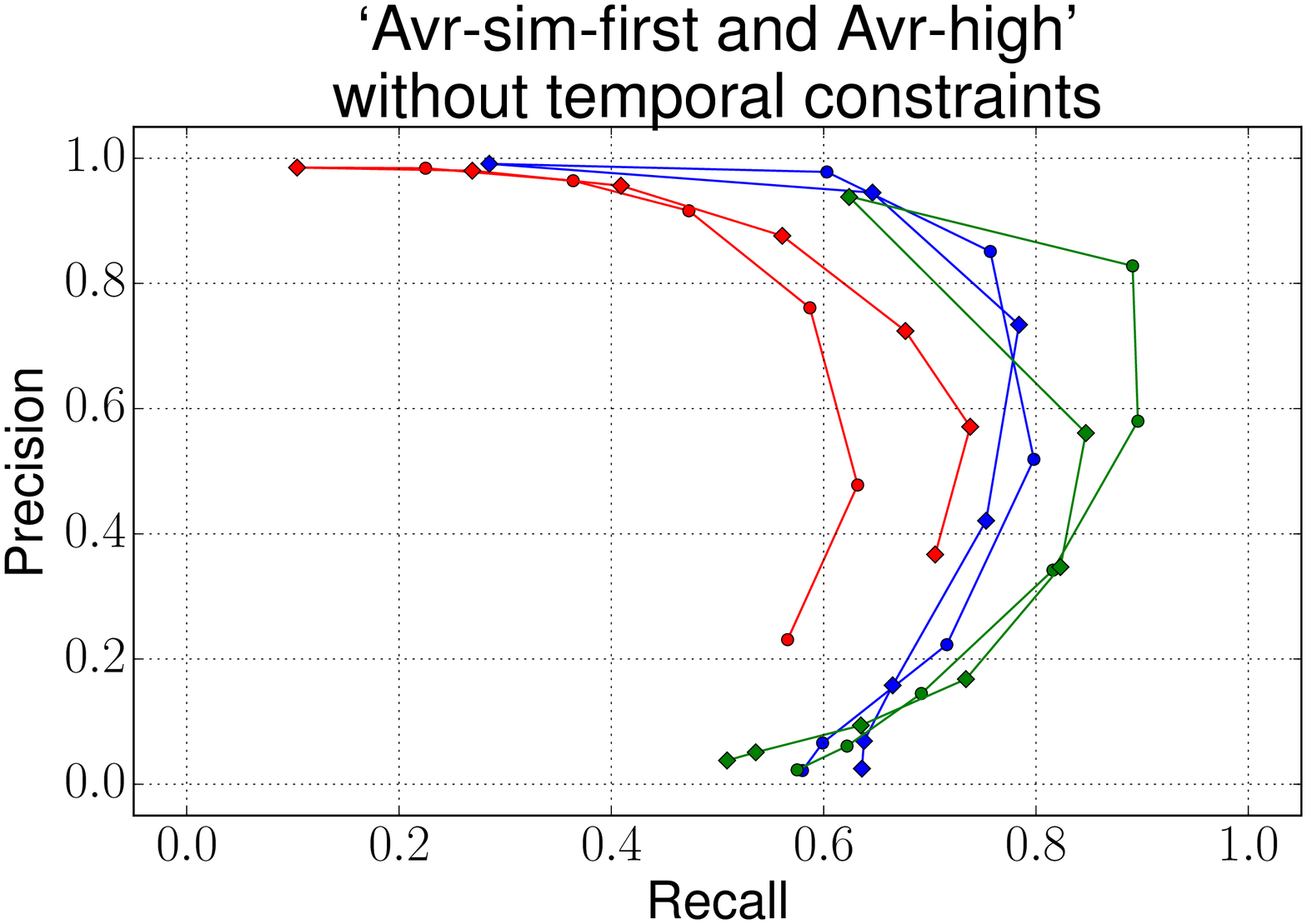}
    \hspace{4mm}
    \includegraphics[width=0.3\textwidth]
      {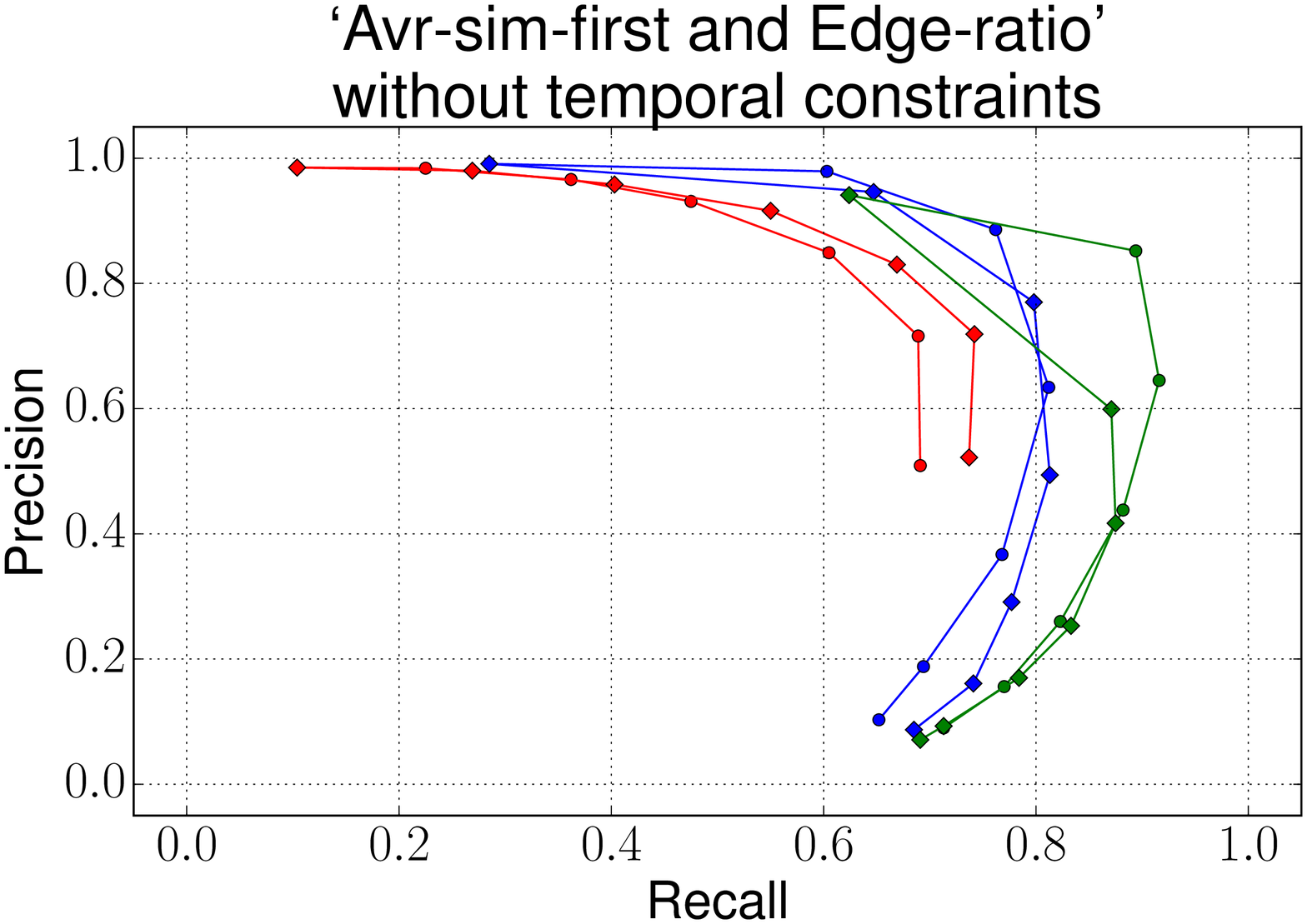}
    \includegraphics[width=0.3\textwidth]
      {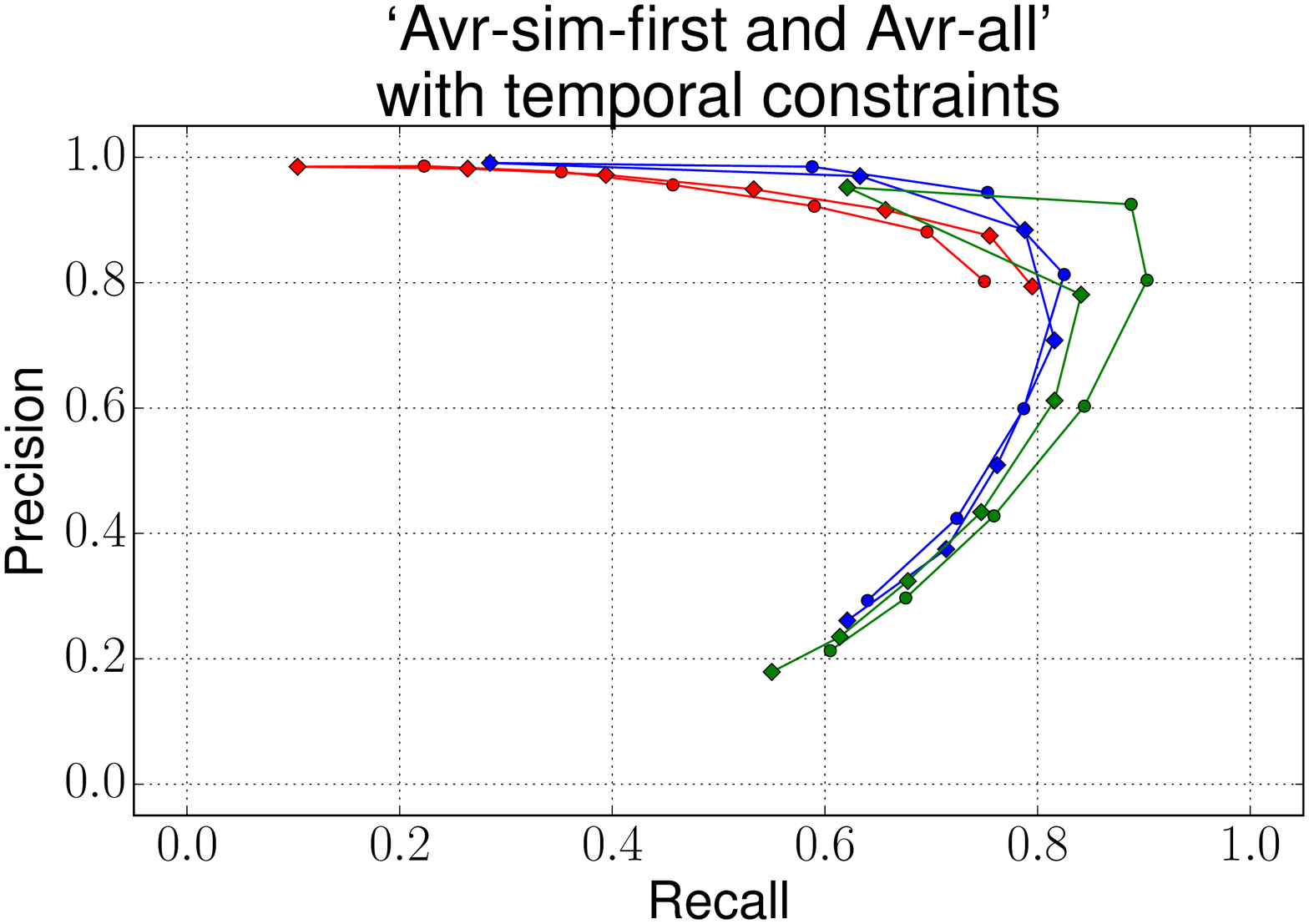}
    \hspace{4mm}
    \includegraphics[width=0.3\textwidth]
      {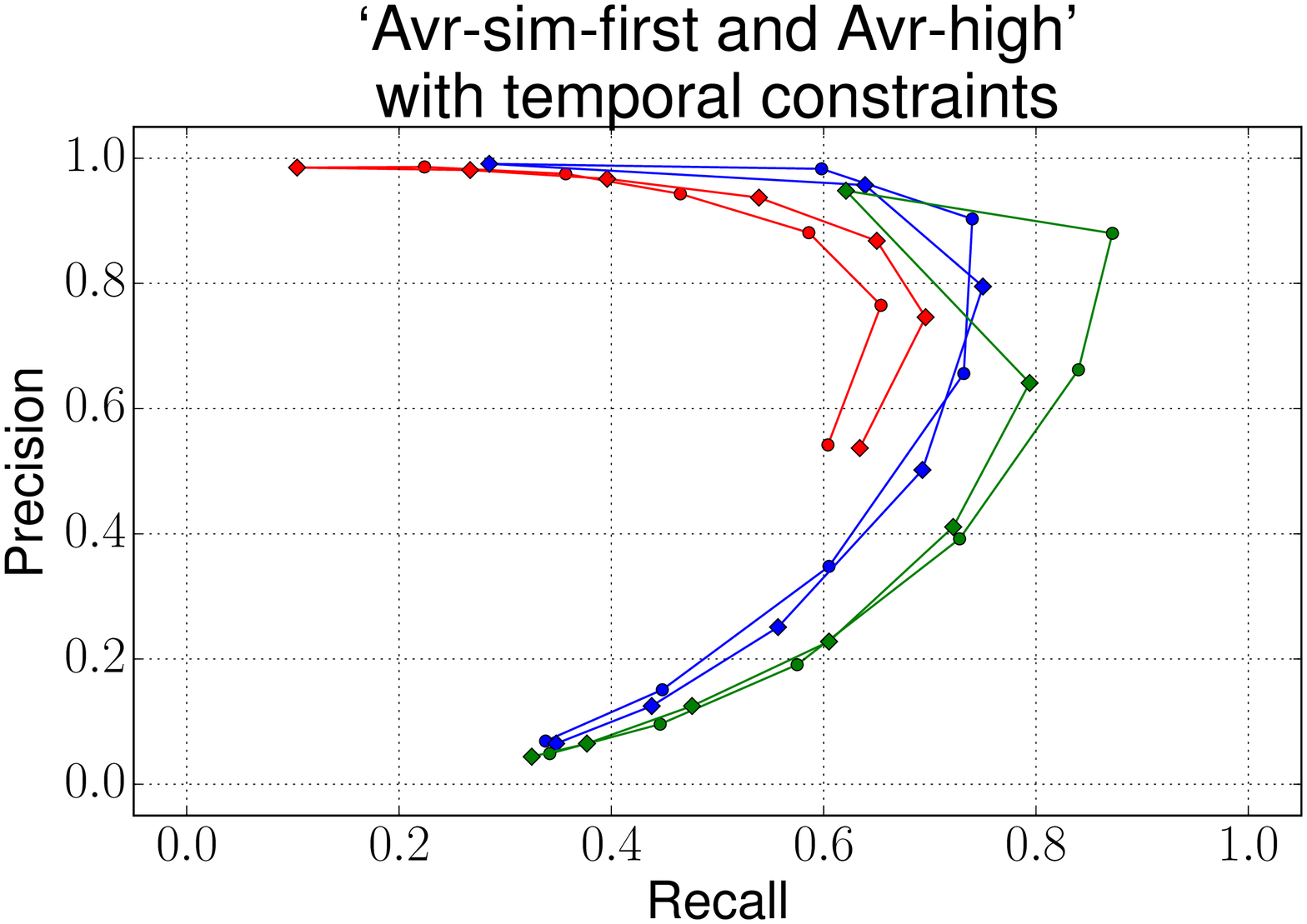}
    \hspace{4mm}
    \includegraphics[width=0.3\textwidth]
      {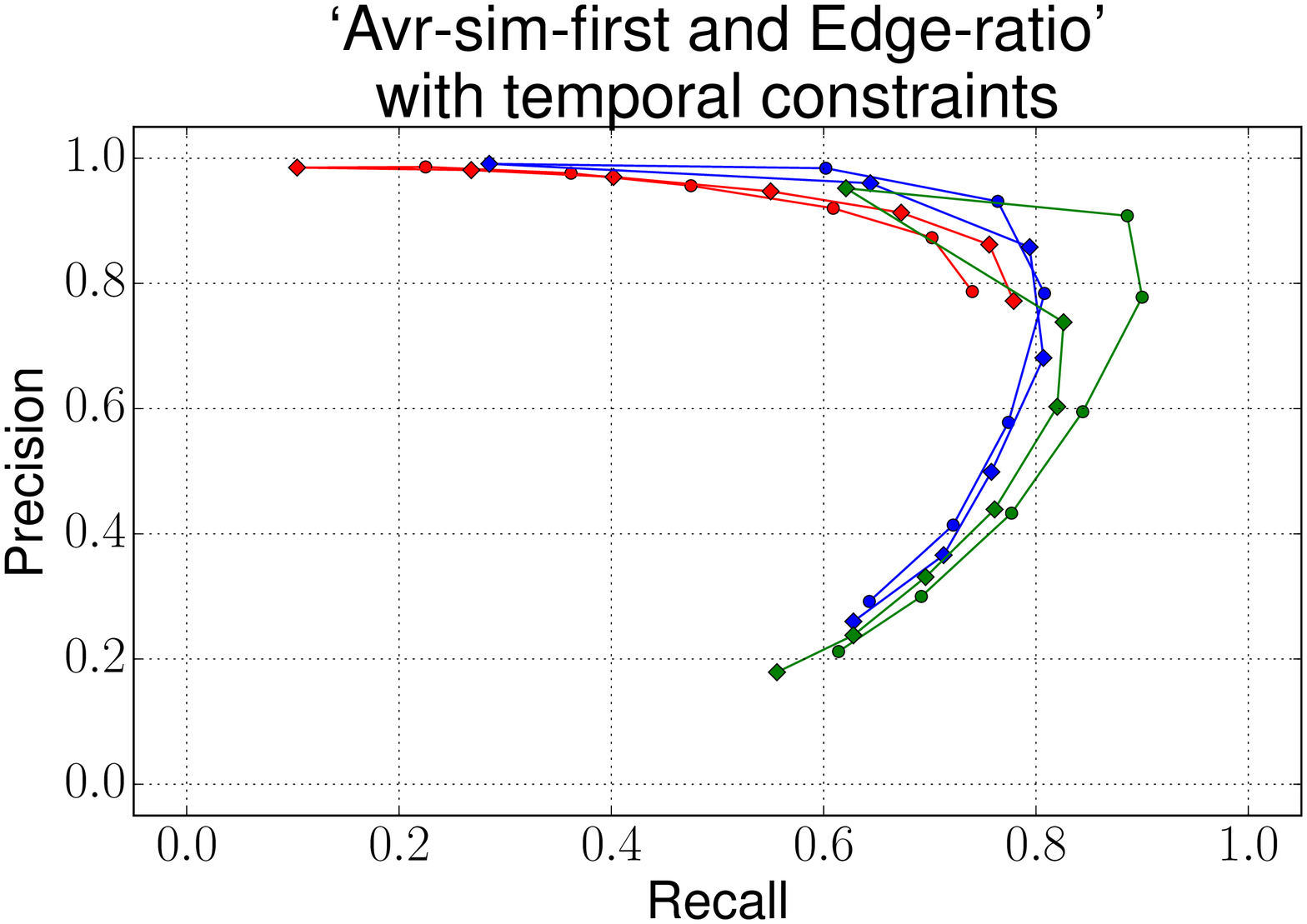}
\caption{Precision-recall results for the temporal star clustering
    approach described in Section~\ref{sec:starcluster} using the
    Avr-sim-first sorting method, the three discussed 
    overlap resolving methods and without
    (top row) and with (bottom row) temporal constraints. Each plot
    shows results for the six similarity graphs described in
    Section~\ref{sec:experiments} (with / without weighted
    similarities and different attributes compared).
    \label{fig:star-cluster-res-1}}
\end{figure*}

\begin{figure*}[t!]
  \centering
    \includegraphics[width=0.3\textwidth]
      {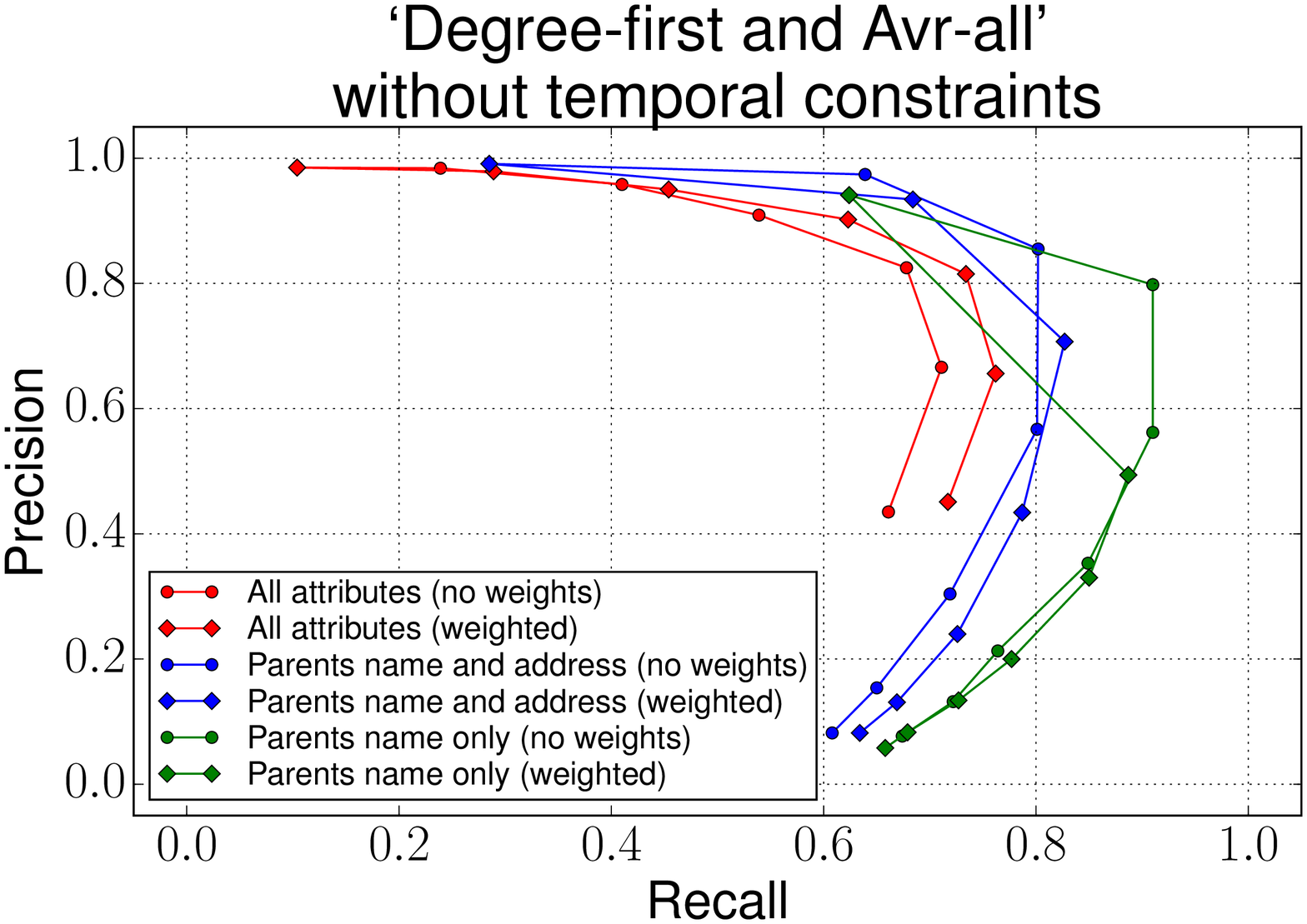}
    \hspace{4mm}
    \includegraphics[width=0.3\textwidth]
      {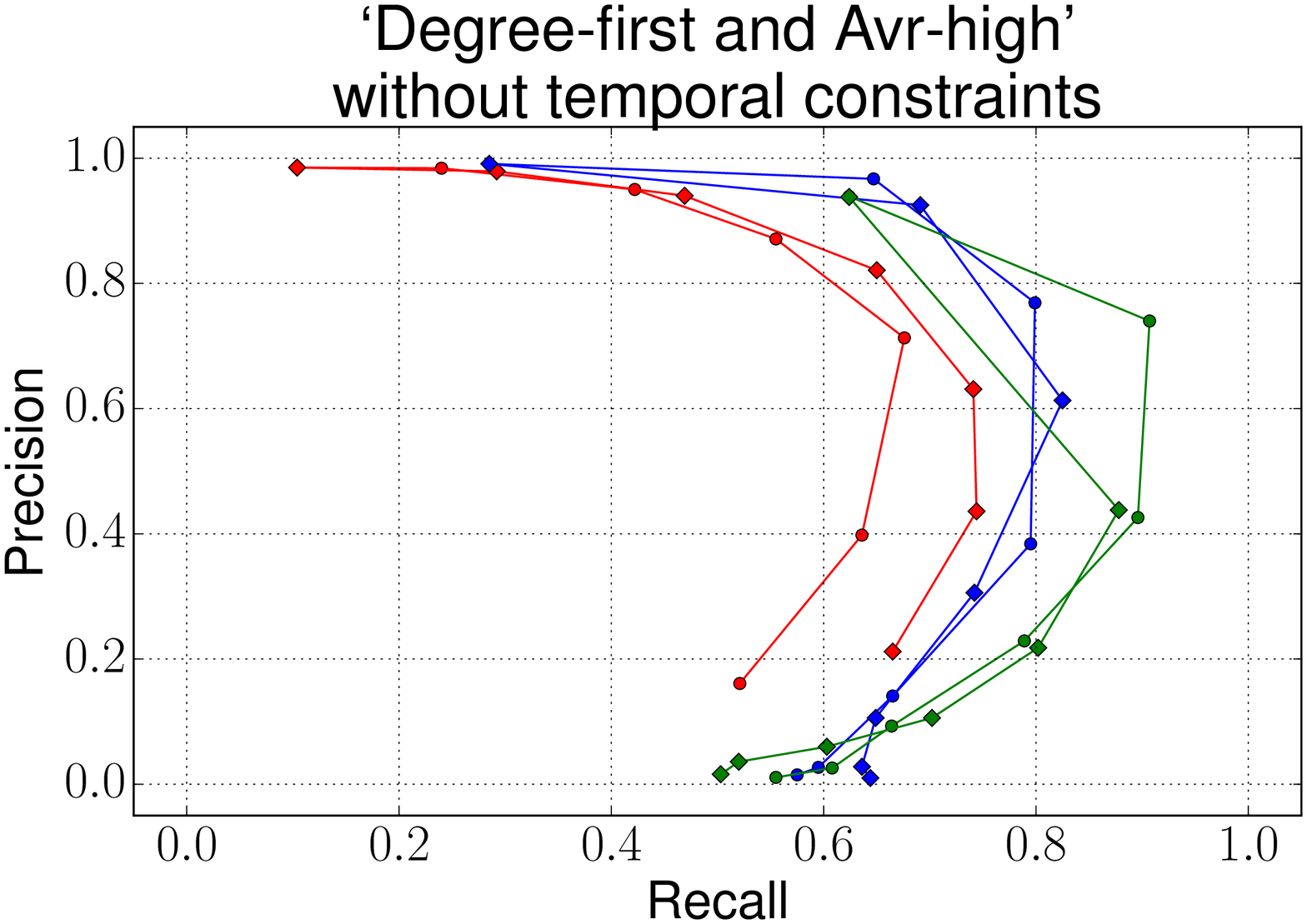}
    \hspace{4mm}
    \includegraphics[width=0.3\textwidth]
      {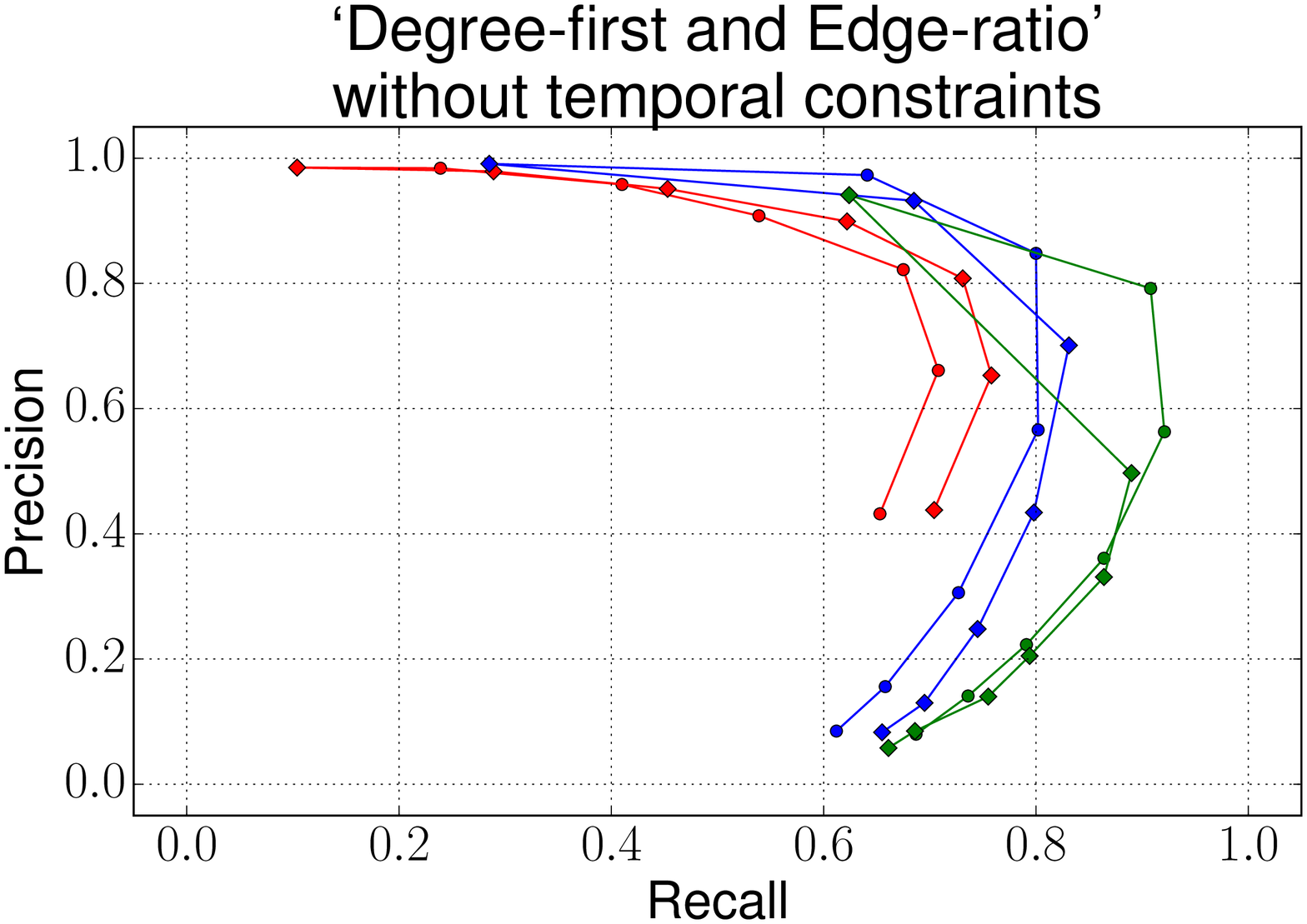}
    \hspace{4mm}
    \includegraphics[width=0.3\textwidth]
      {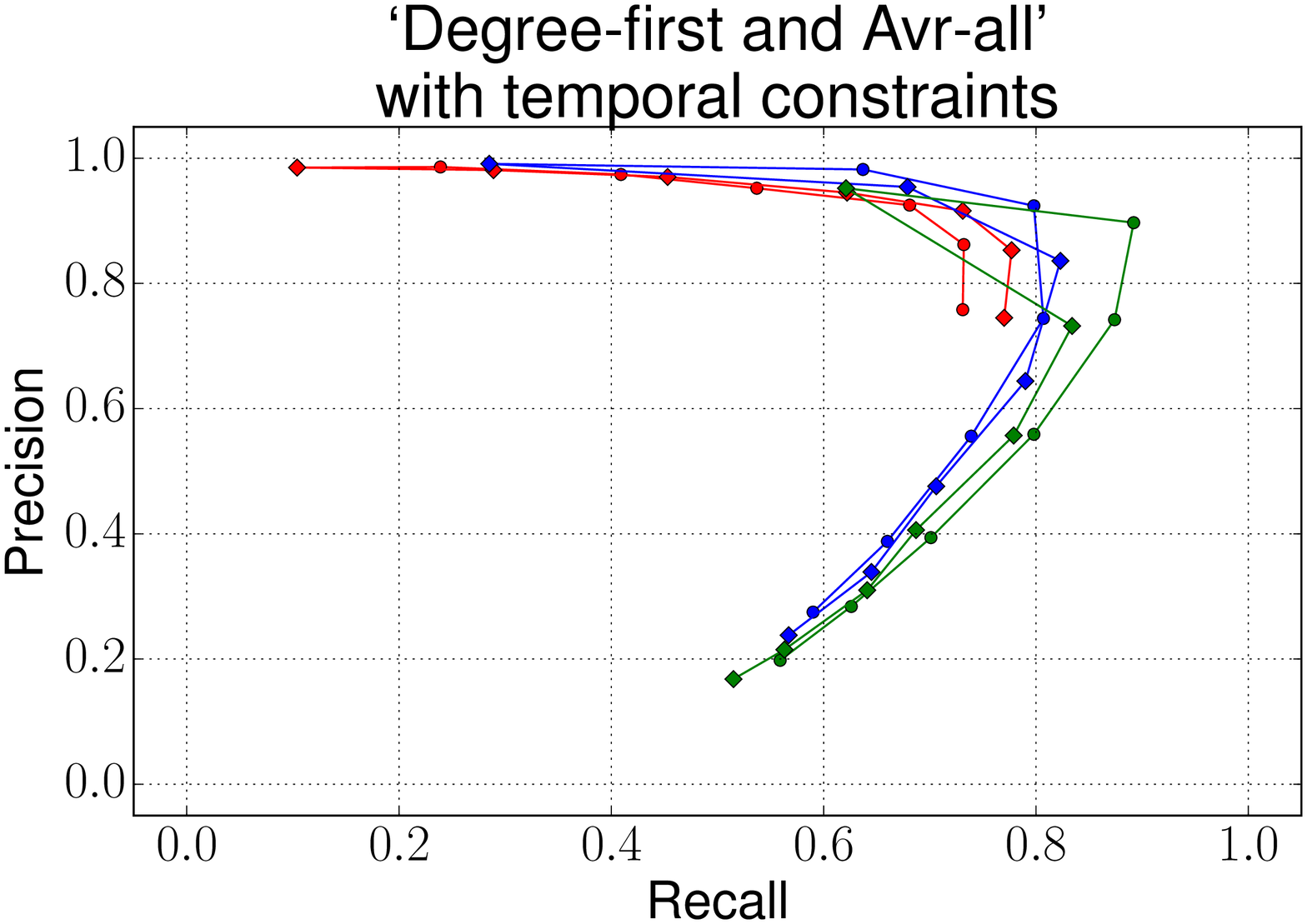}
    \hspace{4mm}
    \includegraphics[width=0.3\textwidth]
      {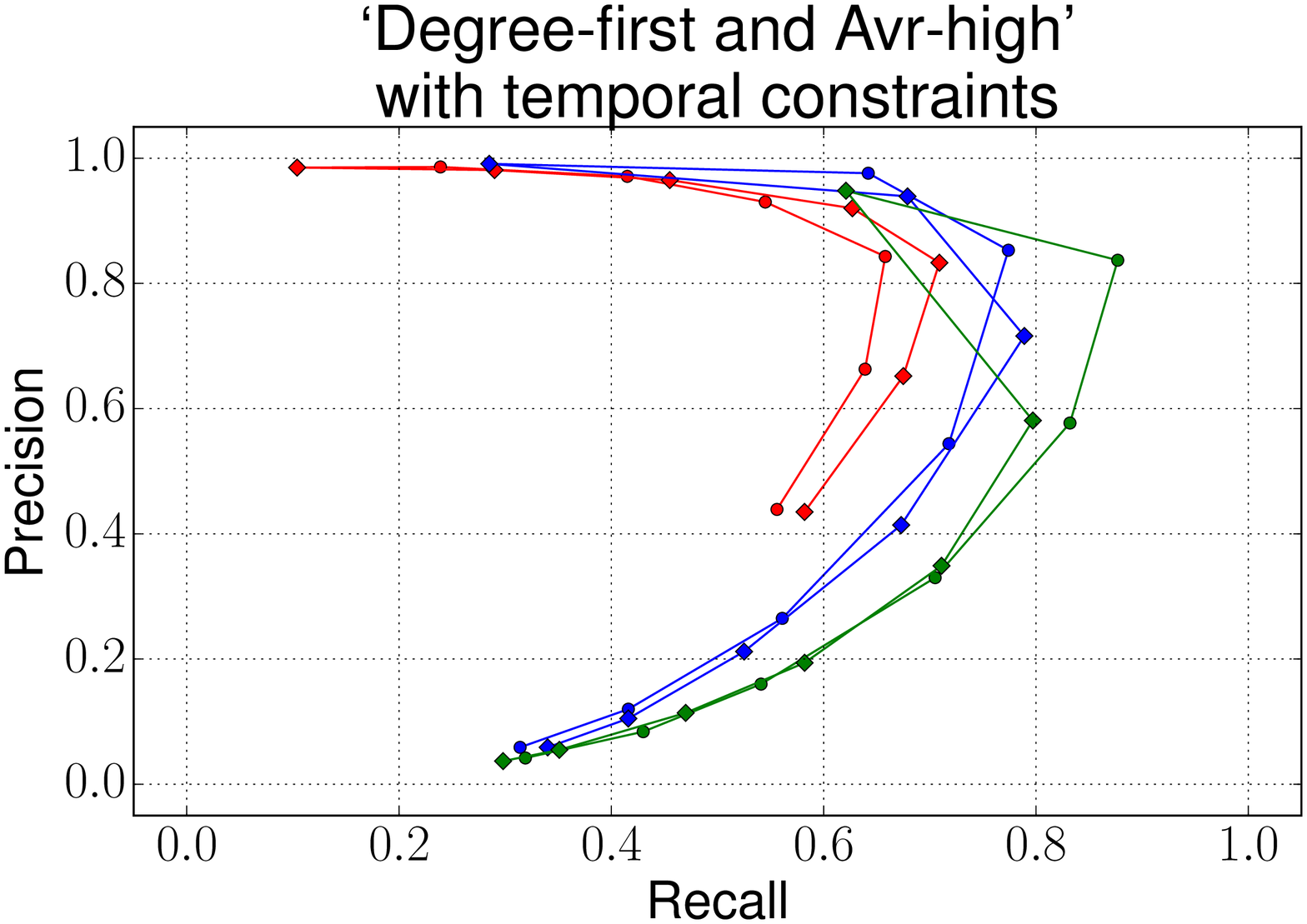}
    \hspace{4mm}
    \includegraphics[width=0.3\textwidth]
      {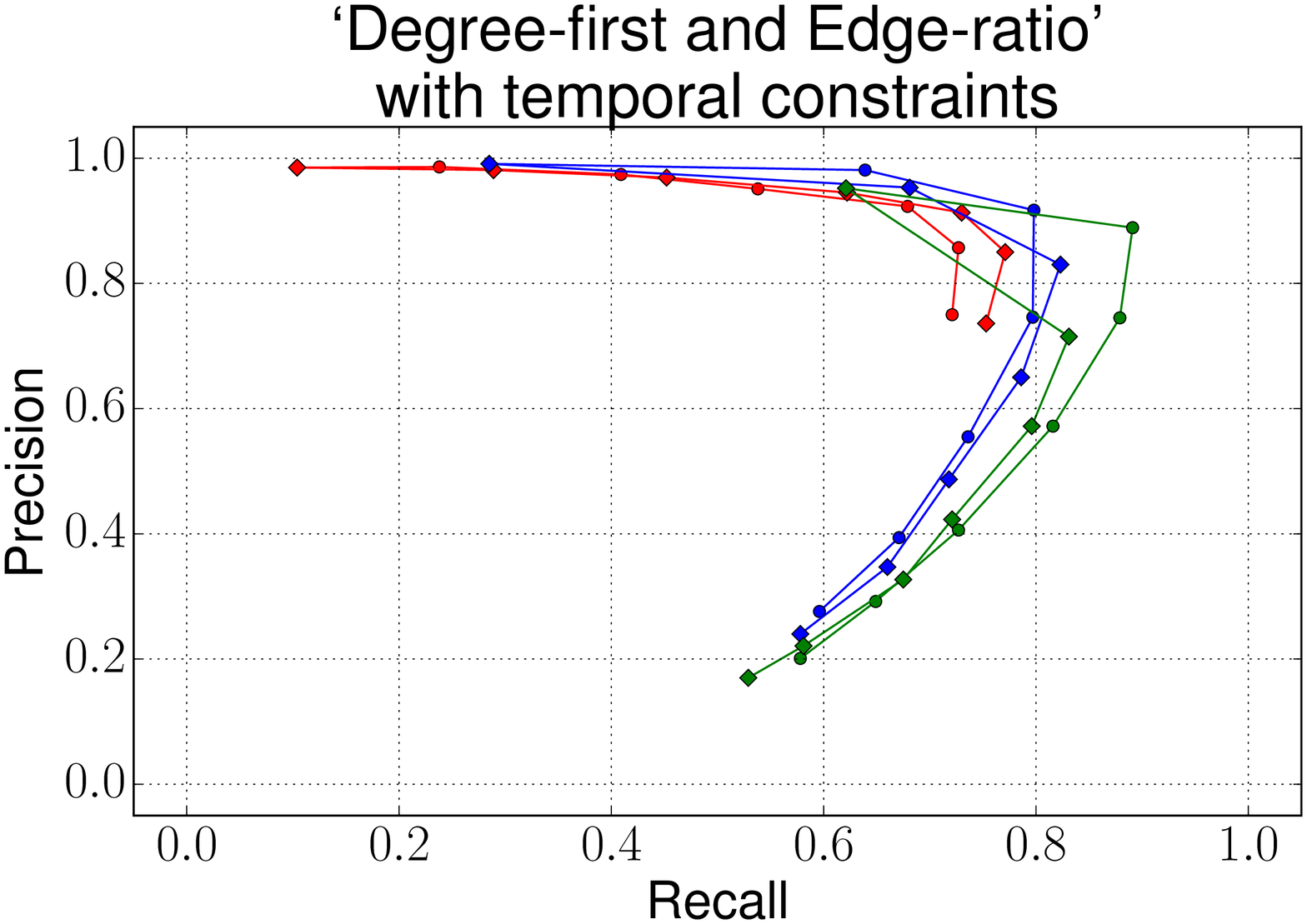}
\caption{Precision-recall results for the temporal star clustering
    approach described in Section~\ref{sec:starcluster} using the
    Degree-first sorting method, the three discussed 
    overlap resolving methods and without
    (top row) and with (bottom row) temporal constraints. Each plot
    shows results for the six similarity graphs described in
    Section~\ref{sec:experiments} (with / without weighted
    similarities and different attributes compared).
    \label{fig:star-cluster-res-2}}
\end{figure*}

\begin{figure*}[t!]
  \centering
    \includegraphics[width=0.3\textwidth]
      {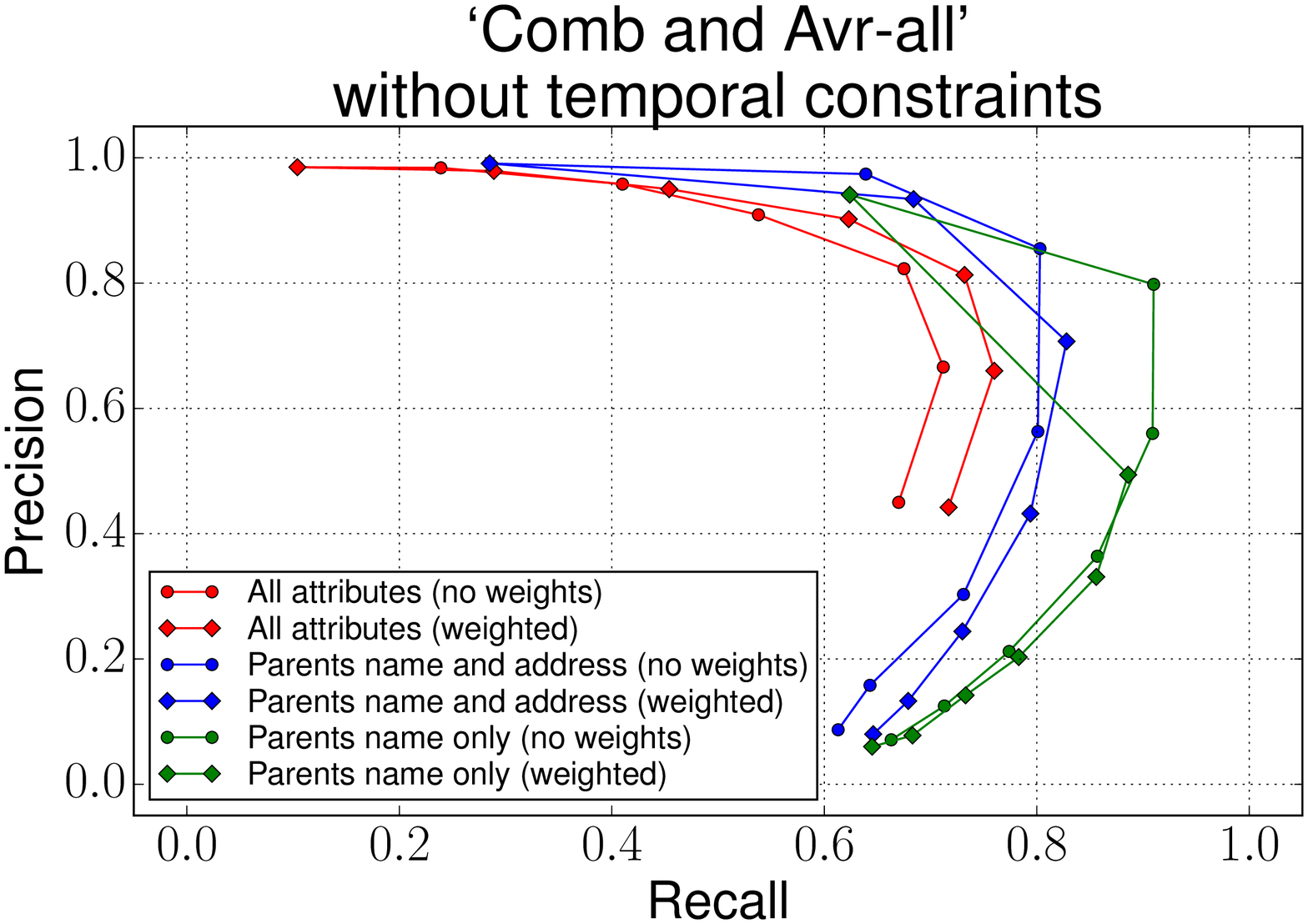}
    \hspace{4mm}
    \includegraphics[width=0.3\textwidth]
      {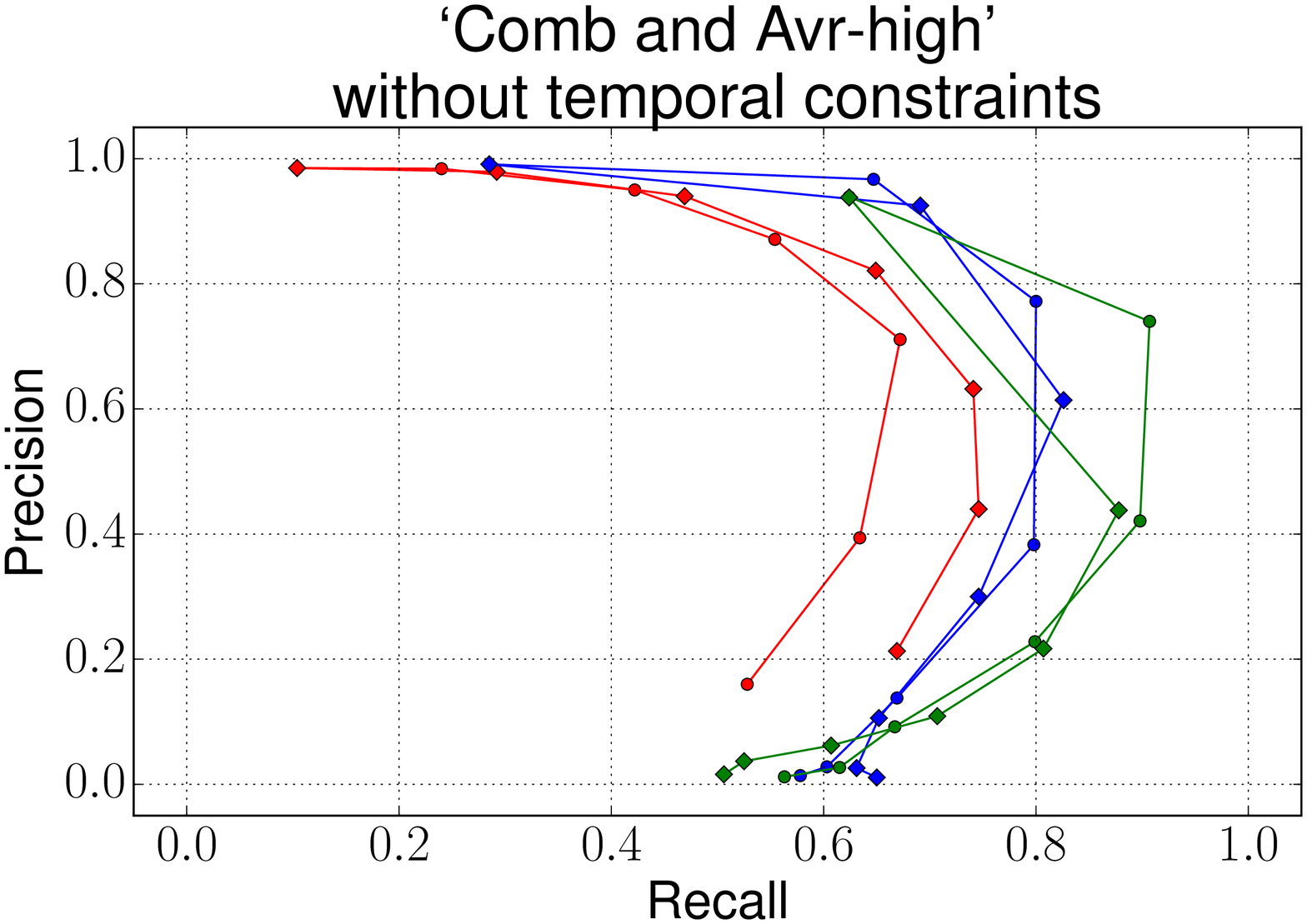}
    \hspace{4mm}
    \includegraphics[width=0.3\textwidth]
      {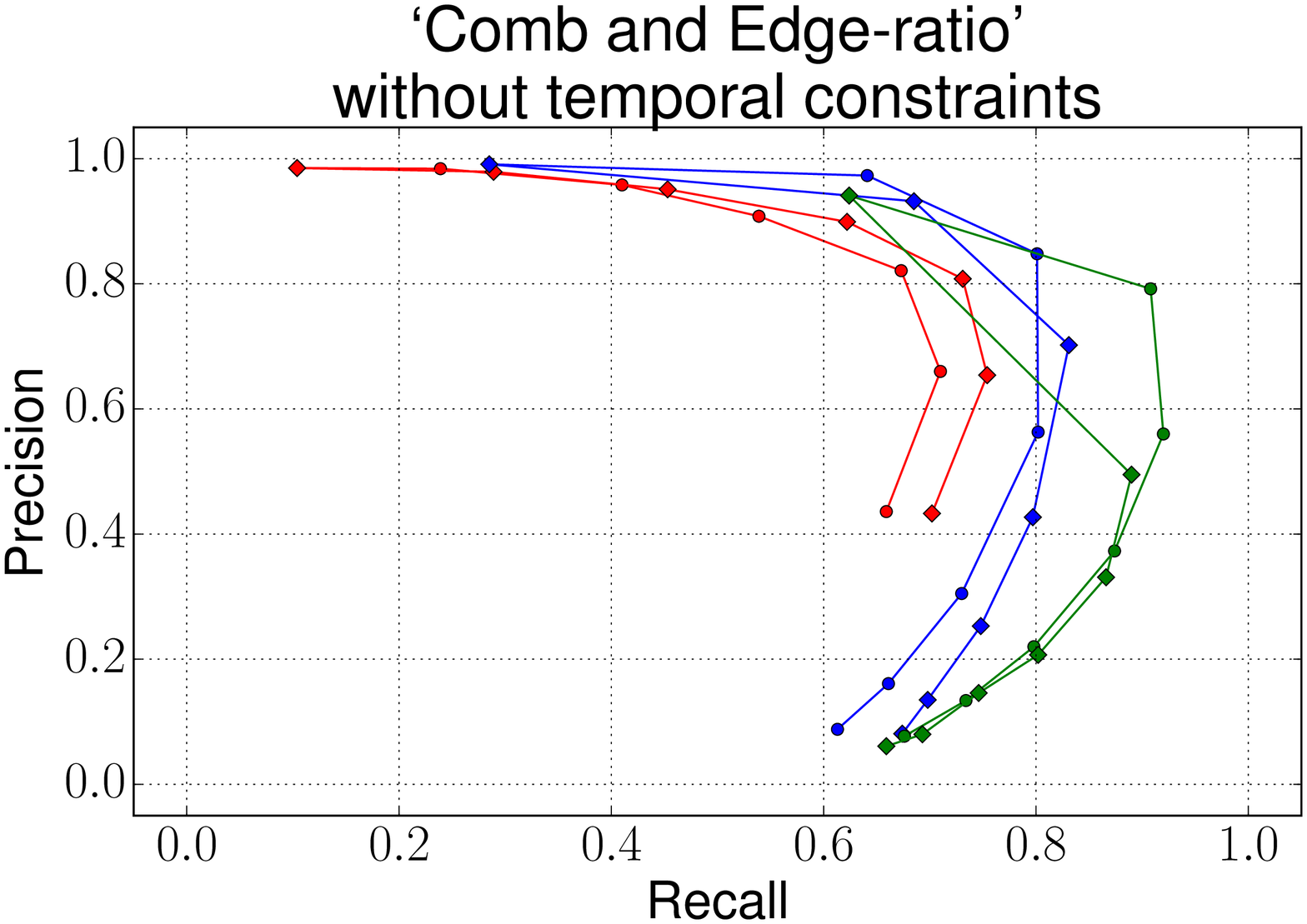}  
    \hspace{4mm}
    \includegraphics[width=0.3\textwidth]
      {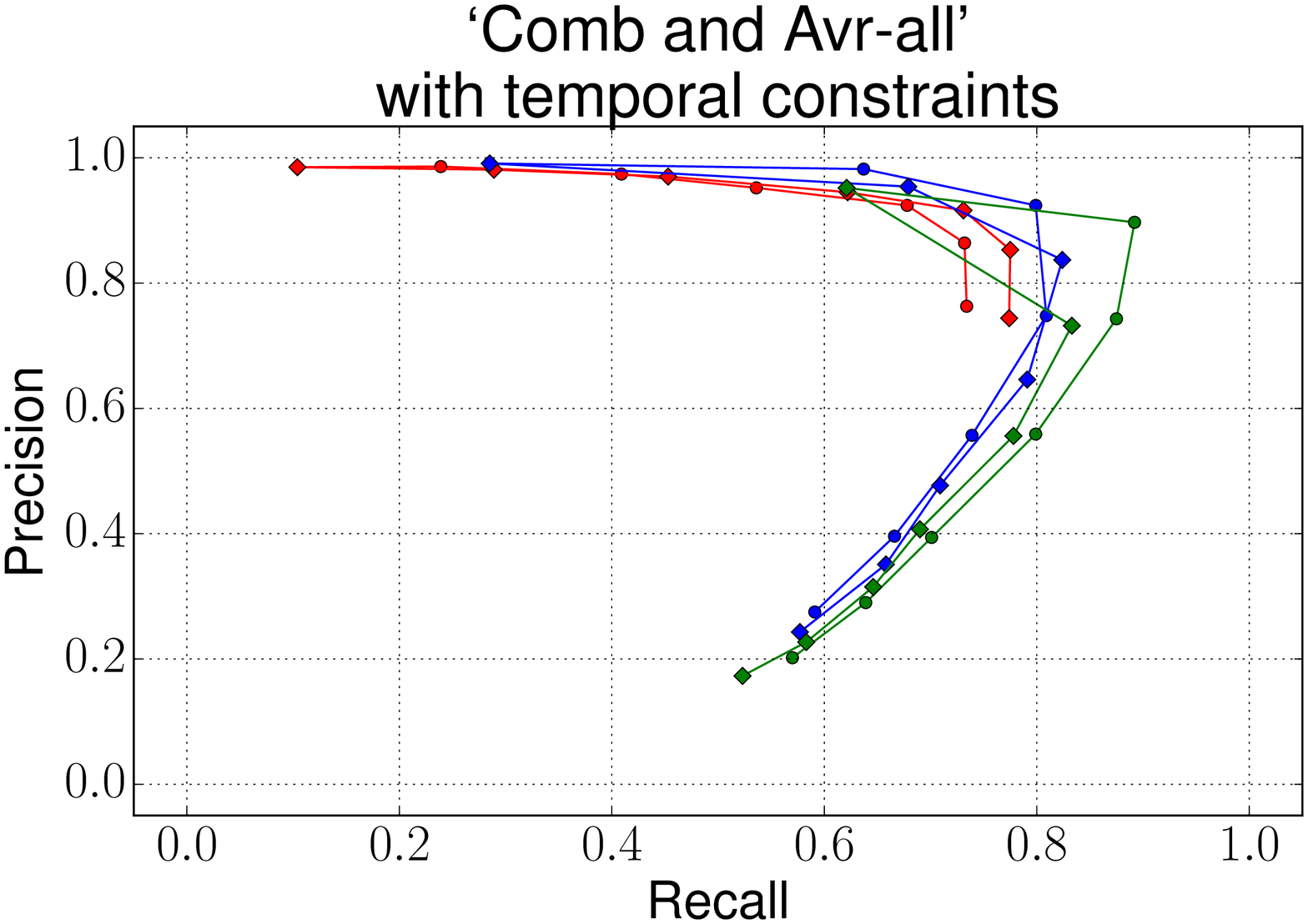}
    \hspace{4mm}
    \includegraphics[width=0.3\textwidth]
      {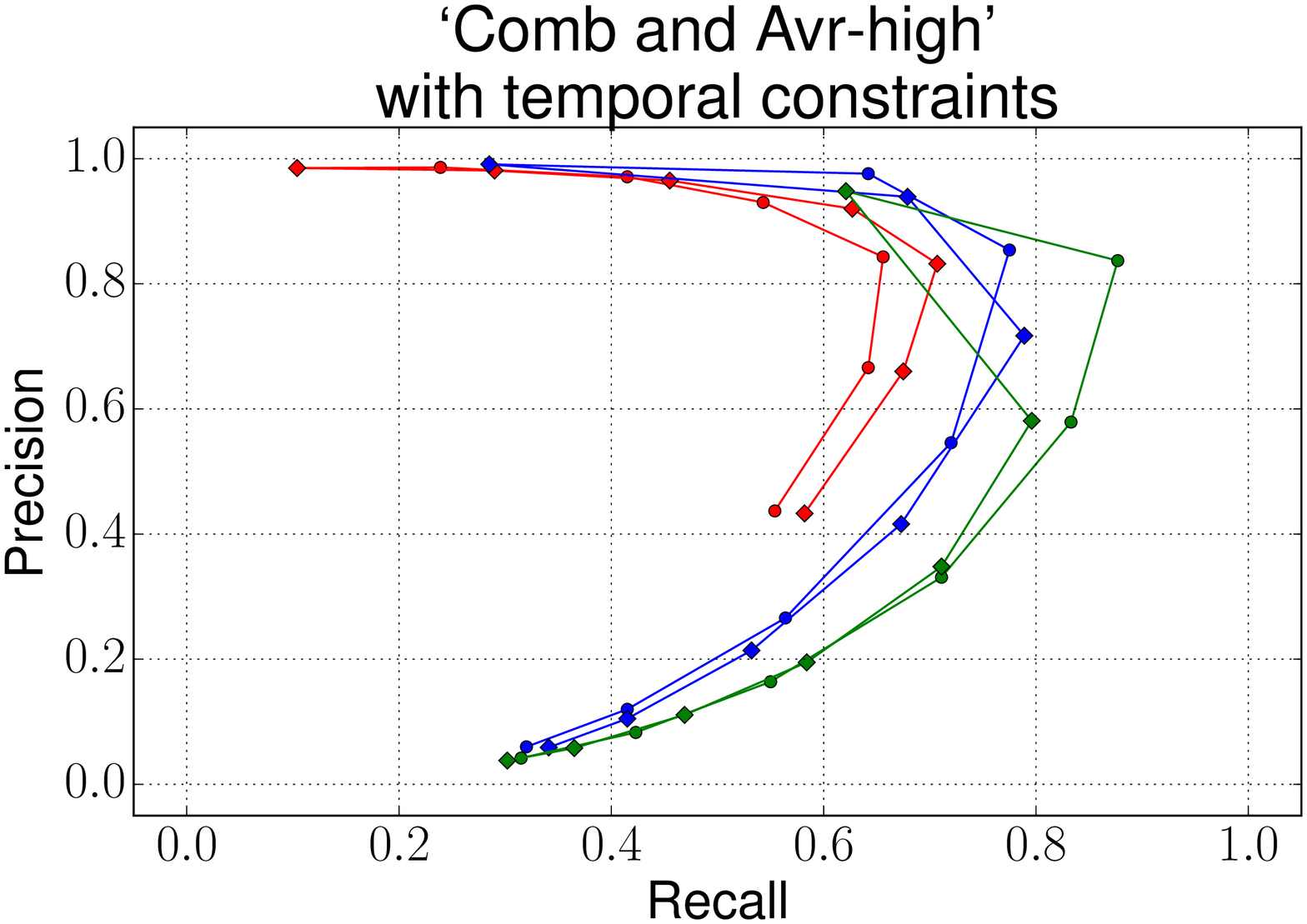}
    \hspace{4mm}
    \includegraphics[width=0.3\textwidth]
      {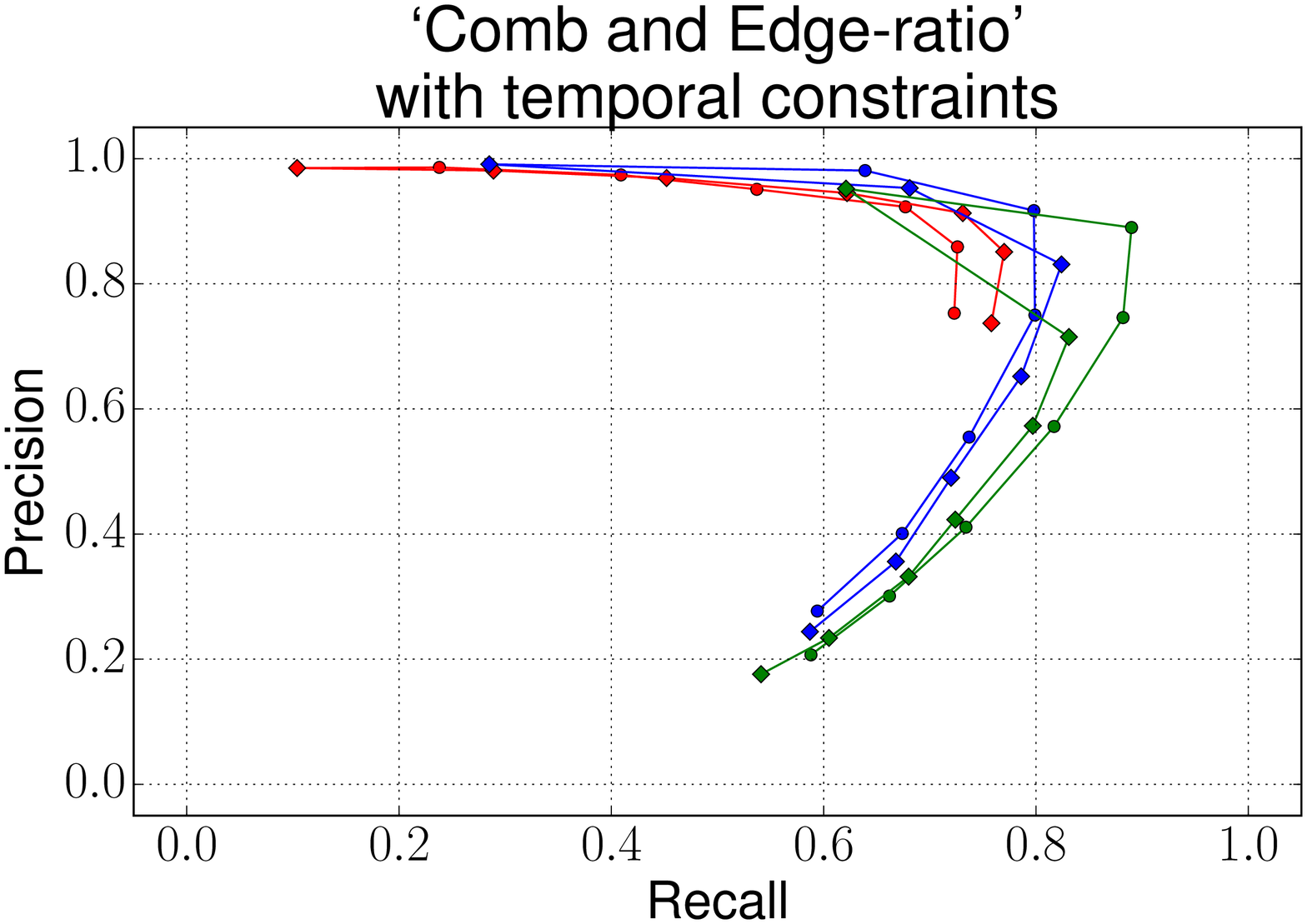}
\caption{Precision-recall results for the temporal star clustering
    approach described in Section~\ref{sec:starcluster} using the
    Comb sorting method, the three discussed 
    overlap resolving methods and without
    (top row) and with (bottom row) temporal constraints. Each plot
    shows results for the six similarity graphs described in
    Section~\ref{sec:experiments} (with / without weighted
    similarities and different attributes compared).
    \label{fig:star-cluster-res-3}}
\end{figure*}

\begin{figure*}[t!]
  \centering
    \includegraphics[width=0.3\textwidth]
      {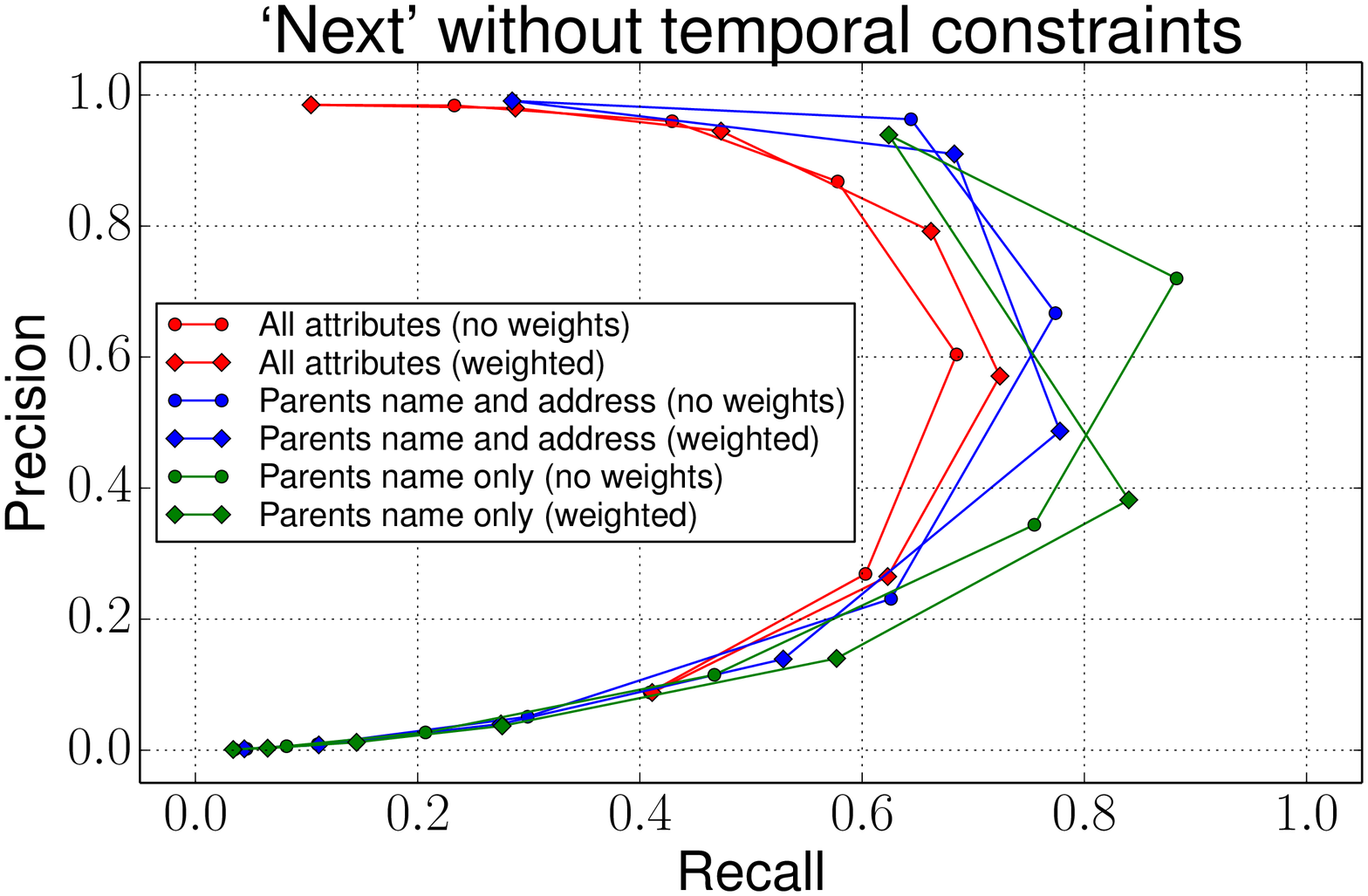}
    \hspace{4mm}
    \includegraphics[width=0.3\textwidth]
      {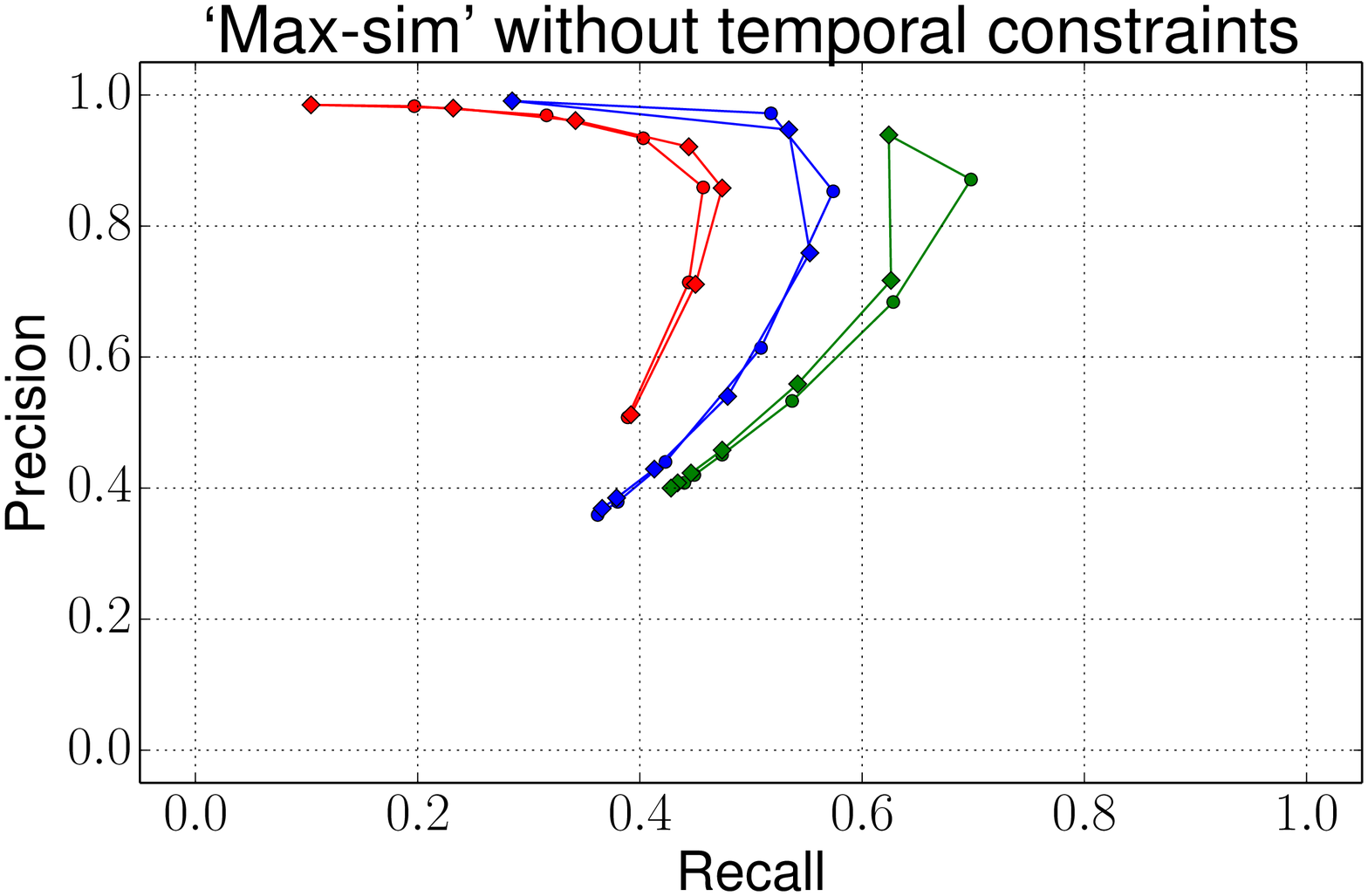}
    \hspace{4mm}
    \includegraphics[width=0.3\textwidth]
      {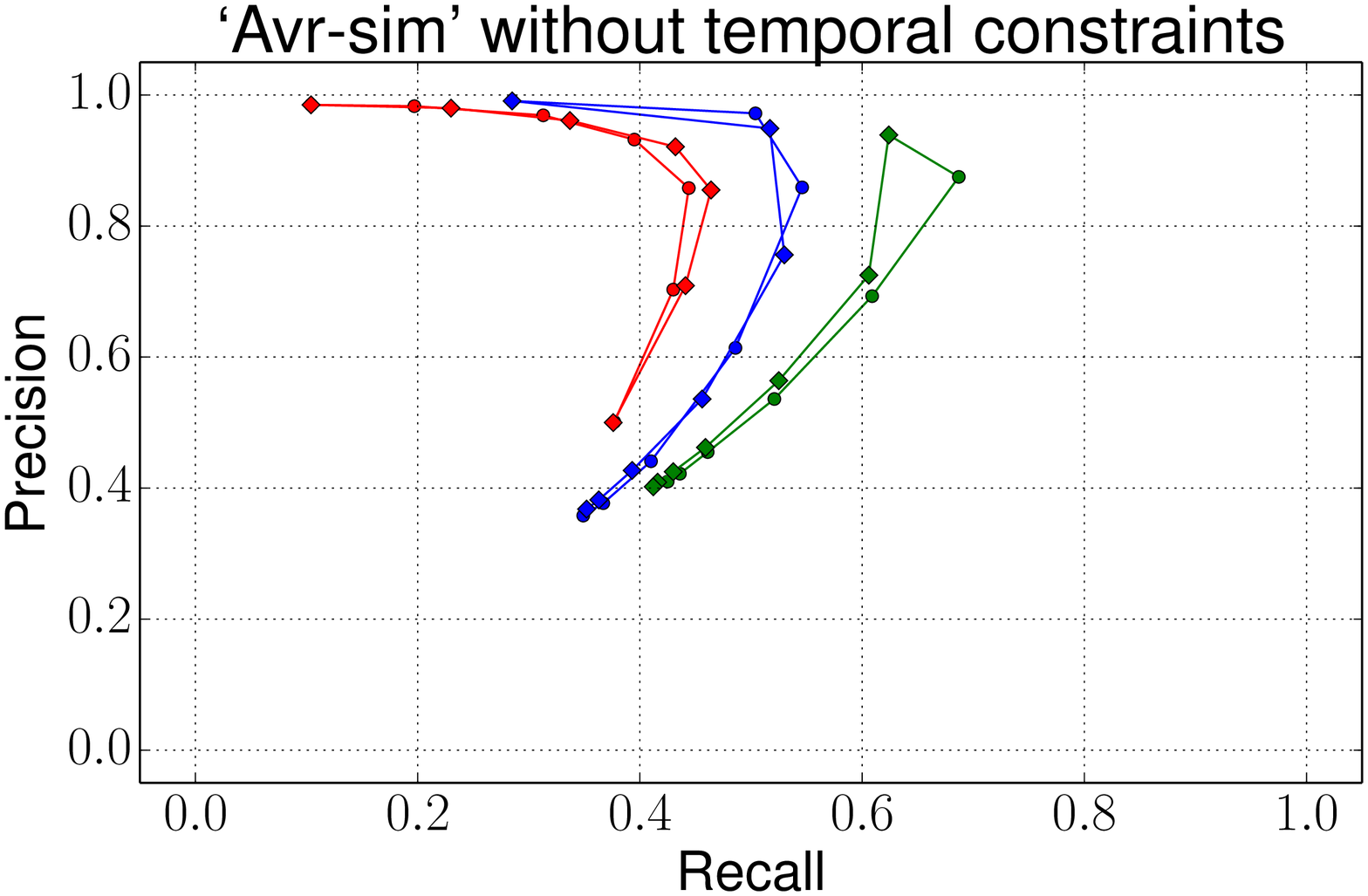}
    \includegraphics[width=0.3\textwidth]
      {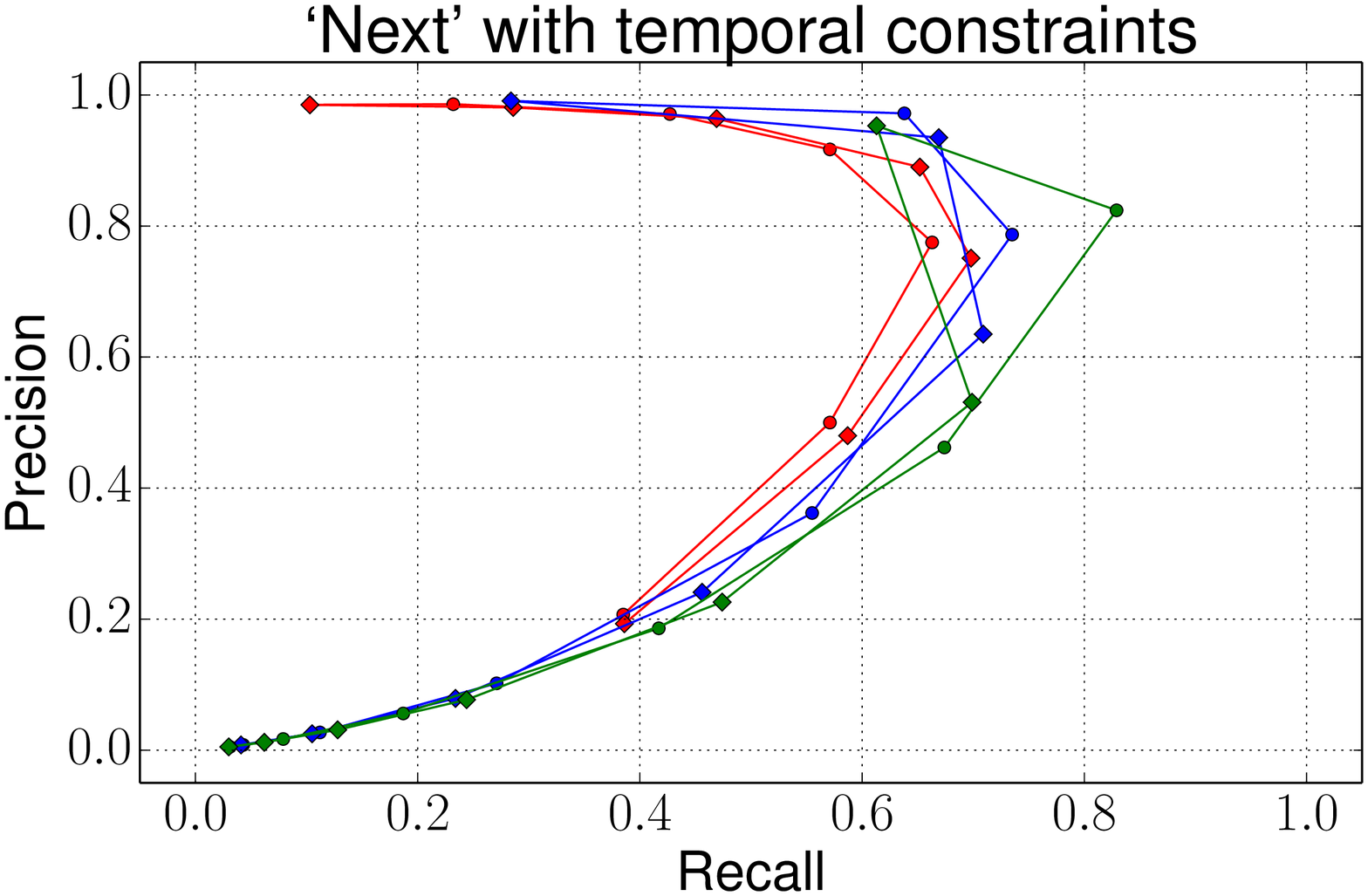}
    \hspace{4mm}
    \includegraphics[width=0.3\textwidth]
      {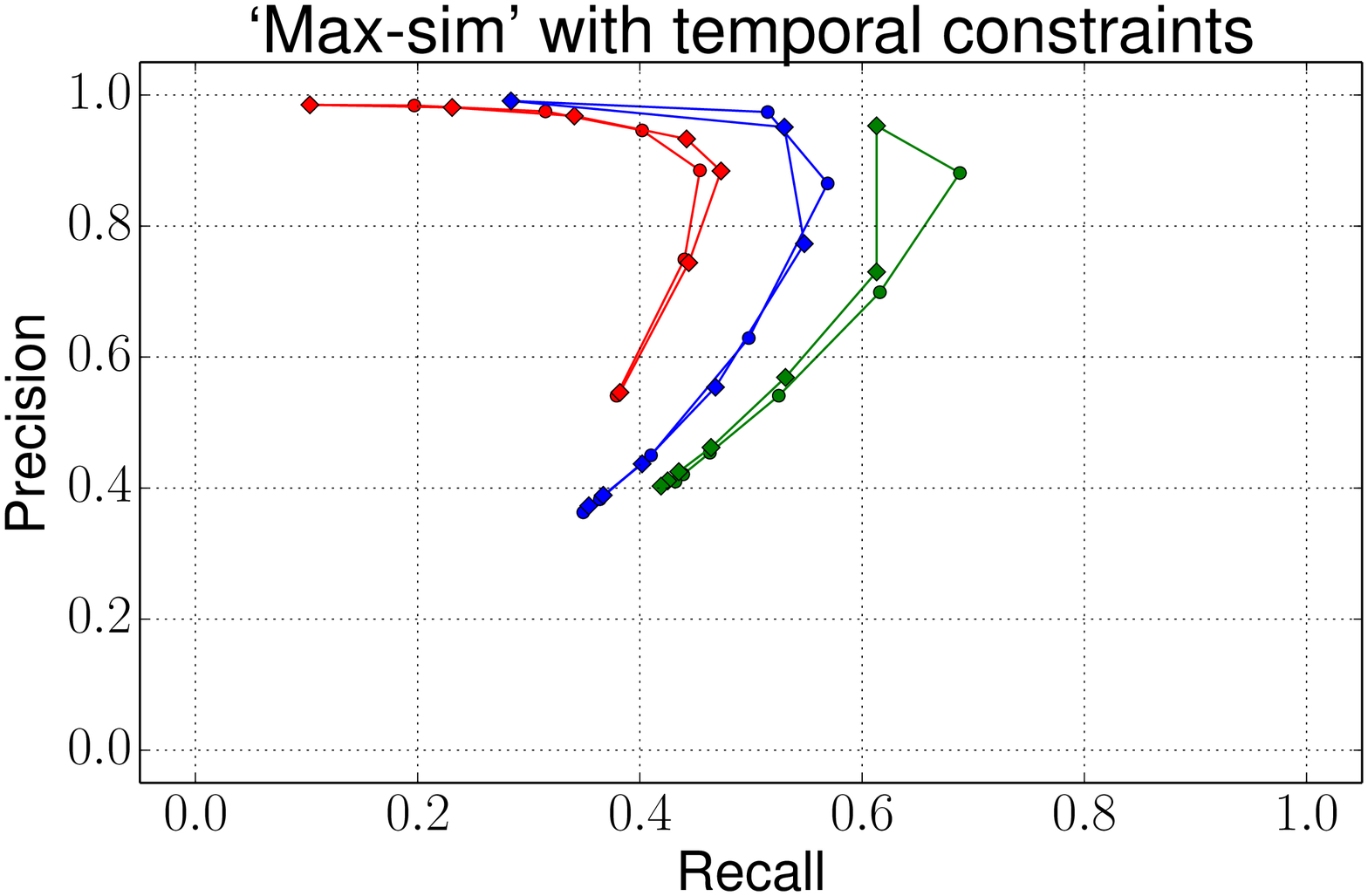}
    \hspace{4mm}
    \includegraphics[width=0.3\textwidth]
      {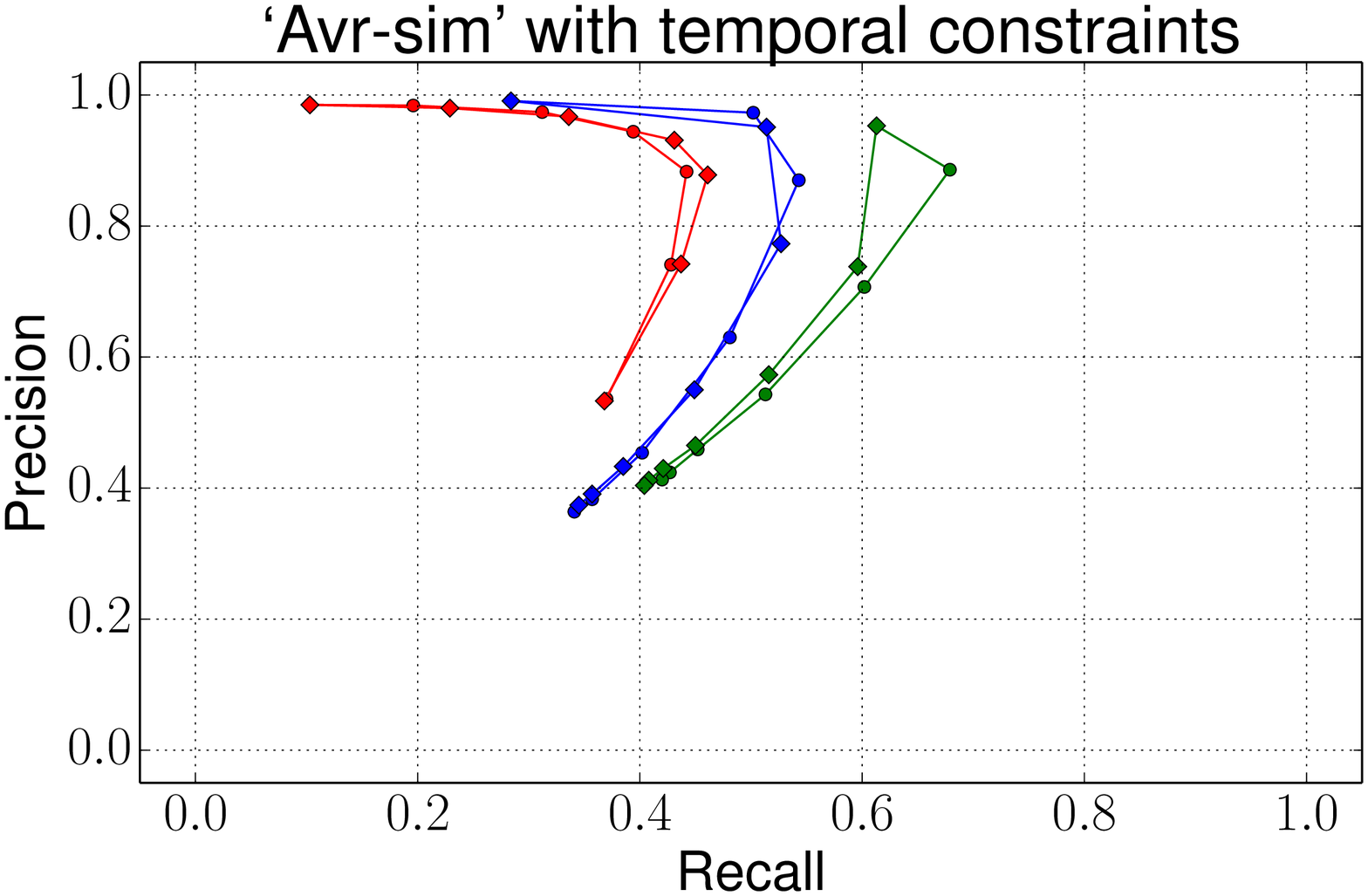}
  \caption{Precision-recall results for the greedy temporal
    clustering approach described in Section~\ref{sec:tempcluster}
    using the three discussed selection methods, and without
    (top row) and with (bottom row) temporal constraints.
    \label{fig:temp-cluster-res}}
\end{figure*}

For the clustering approaches described in
Sections~\ref{sec:starcluster} and~\ref{sec:tempcluster}, we
evaluate the three sorting and resolving methods for star clustering,
and the three selection methods for greedy temporal clustering.
We show the final clustering results obtained as precision-recall
curves in Figures~\ref{fig:star-cluster-res-1} to 
~\ref{fig:temp-cluster-res} where we changed the value of the
minimum similarity threshold to include pair-wise similarities
(i.e.\ edges) in the graph $\mathbf{G}$ from $1.0$ to $0.7$ in
$0.05$ steps.

These rather unusual looking precision-recall curves need some
explanation. When the minimum similarity threshold $s_{min}$ used to
generate the pair-wise graph $\mathbf{G}$ is lowered, more false
matches are included as edges into $\mathbf{G}$, thus reducing the
precision as expected. However, recall seems to have an inverse
relationship with $s_{min}$ up-to a certain point (recall increases
while $s_{min}$ is decreased) and then recall decreases with
$s_{min}$. We believe that this behaviour is caused by the greedy
nature of the algorithms and the skewness of the attribute value
distribution. When $s_{min}$ is too high (such as 1.0), many
true-matches which are not exact matches (due to mistakes in data
transposition, etc.) get dropped, leading to lower recall. When
$s_{min}$ is slightly more lenient (such as 0.95 or 0.9), recall
improves since more of the true-matches with slight spelling mistakes
are included into clusters and are therefore matched. However, when
$s_{min}$ is further lowered, the number of high similarity
non-matches increases (due to skewness of the distribution) and
these non-matches will be clustered incorrectly. This is caused by
the greedy nature of both clustering algorithms, where after an
incorrect node is selected as the next best node the actual true
matches are never offered a chance to be clustered together. This
behaviour is mostly accentuated when only parent names are used to
calculate the similarities between certificates which is because
the distribution of parent names is the most skewed.

As Figures~\ref{fig:star-cluster-res-1} to 
~\ref{fig:temp-cluster-res} show, when temporal constraints are
included in the clustering phase then precision generally increases
considerably while recall only decreases little. The overall best
performing approach (with and without temporal constraints) 
was using unweighted similarities of only parent
names, with Avr-sim-first as the sorting method and Avr-all as the 
overlap resolving method. Furthermore, the similarity threshold 
value achieving the best results was 0.95.
The overall highest precision and recall results without
temporal constraints were $0.877$ and $0.897$, while when applying
temporal constraints they were $0.925$ and $0.888$, respectively.

The result plots also show that overall star clustering achieves
better results with regard to recall than the temporal greedy
technique, however the similarity based selection methods for
temporal greedy clustering achieve overall higher minimum
precision results.


\section{Conclusions and Future Work}
\label{sec:conclusions}

In this work-in-progress paper we have developed and evaluated two
clustering approaches for linking birth certificates in the context
of historical record linkage. Both algorithms are based on a graph
that represents the similarities calculated between individual
birth certificates. We have evaluated six approaches how this
graph is generated based on comparing different attribute combinations
in a weighted or unweighted fashion, and how the characteristics of
this graph affect the final clustering outcomes. Our experimental
evaluation on a real Scottish data set have shown that incorporating
temporal constraints (when a woman can give birth or not) can
improve the quality of the final linked data set.

As future work we aim to improve our proposed greedy temporal
clustering algorithm as well as temporal star clustering to obtain
better linkage results. We aim to investigate why certain birth
certificates are not linked (missed true matches, lowering recall)
while others are falsely linked (wrong matches, lowering precision).
We then aim to expand our graph-based clustering techniques to also
incorporate links across birth, marriage, death, and census
certificates by generating a single large similarity graph where
nodes represent certificates and edges, the similarities between them,
and where edges can be of different types~\cite{Chr16}. Such a graph
will not only allow temporal constraints to be considered but also
gender and role-type specific constraints~\cite{Chr16,Chr17}.
We plan to model temporal aspects of how the records about a certain
individual will occur in historical population databases. Our
ultimate aim is to develop unsupervised techniques for the accurate
and efficient linkage of large and complex historical population
databases in order to provide researchers in areas such as health
and the social sciences with high quality longitudinal data sets.


\section*{Acknowledgements}

\begin{small}
This work was supported by ESRC grants ES/K00574X/2 \emph{Digitising
Scotland} and ES/L007487/1 \emph{Administrative Data Research Centre
-- Scotland}. We like to thank Alice Reid of the University of
Cambridge and her colleagues Ros Davies and Eilidh Garrett for their
work on the Isle of Skye database, and their helpful advice on
historical Scottish demography. This work was partially funded by the
Australian Research Council under DP130101801.
\end{small}


\bibliographystyle{ACM-Reference-Format}
\bibliography{paper}

\end{document}